\documentclass[journal]{IEEEtran}

\usepackage{amsmath}
\usepackage{amssymb}

\usepackage{graphicx}
\usepackage{epstopdf}
\usepackage{amsmath}
\usepackage{array}
\newcommand{\PreserveBackslash}[1]{\let\temp=\\#1\let\\=\temp}
\newcolumntype{C}[1]{>{\PreserveBackslash\centering}p{#1}}
\newcolumntype{R}[1]{>{\PreserveBackslash\raggedleft}p{#1}}
\newcolumntype{L}[1]{>{\PreserveBackslash\raggedright}p{#1}}
\usepackage[usenames]{color}
\usepackage{colortbl,booktabs}
\usepackage{tabularx}
\usepackage{multicol}
\usepackage{booktabs}
\usepackage[Symbol]{upgreek}
\usepackage{bm}
\usepackage{cite}
\usepackage{balance}
\usepackage[boxed]{algorithm2e}
\usepackage{mathrsfs}
\usepackage{url}
\usepackage{threeparttable}
\usepackage{amsmath}
\usepackage{dcolumn}
\usepackage{multirow}
\usepackage{booktabs}
\usepackage{amsfonts}
\usepackage{algorithmic} 

\newcommand{\tabincell}[2]{\begin{tabular}{@{}#1@{}}#2\end{tabular}}
\begin{document}
	
	\title{Channel Estimation for Orthogonal Time Frequency Space (OTFS) Massive MIMO \vspace{-0mm}}
	
	\author{
		Wenqian Shen,~\IEEEmembership{Student Member,~IEEE}, Linglong Dai,~\IEEEmembership{Senior Member,~IEEE}, Jianping An,~\IEEEmembership{Member,~IEEE}, Pingzhi Fan,~\IEEEmembership{Fellow,~IEEE}, and Robert W. Heath, Jr.,~\IEEEmembership{Fellow,~IEEE}
		\thanks{
			W. Shen and J. An are with the School of Information and Electronics, Beijing Institute of Technology, Beijing 100081, China (e-mails: wshen@bit.edu.cn, an@bit.edu.cn).
			
			L. Dai is with the Department of Electronic Engineering, Tsinghua University, Beijing 100084, China (e-mail: daill@tsinghua.edu.cn). 
			
			P. Fan is with the Institute of Mobile Communications, Southwest Jiaotong University, Chengdu 610031, China (e-mail: pzfan@swjtu.edu.cn).
			
			R. W. Heath Jr. is with the Department of Electrical and Computer Engineering, The University of Texas at Austin, Austin, TX 78712-1687, USA (e-mail: rheath@utexas.edu). R. W. Heath Jr. is also on the Technical Advisory Board of Cohere Technologies, which developed OTFS. The terms of this arrangement have been reviewed and approved by the University of Texas at Austin in accordance with its policy on objectivity in research.
			
		}
		\vspace{-0
			mm}}
	
	\maketitle
	
	\begin{abstract}
		Orthogonal time frequency space (OTFS) modulation outperforms orthogonal frequency division multiplexing (OFDM) in high-mobility scenarios. One challenge for OTFS massive MIMO is downlink channel estimation due to the large number of base station antennas. In this paper, we propose a 3D structured orthogonal matching pursuit algorithm based channel estimation technique to solve this problem. First, we show that the OTFS MIMO channel exhibits 3D structured sparsity: normal sparsity along the delay dimension, block sparsity along the Doppler dimension, and burst sparsity along the angle dimension. Based on the 3D structured channel sparsity, we then formulate the downlink channel estimation problem as a sparse signal recovery problem. Simulation results show that the proposed algorithm can achieve accurate channel state information with low pilot overhead.
		
	\end{abstract}
	
	
	\IEEEpeerreviewmaketitle
	
	\section{Introduction}\label{S1}
	One goal of future wireless communications (the emerging 5G or beyond 5G) is to support reliable communications in high-mobility scenarios, such as on high-speed railways with a speed of up to 500 km/h \cite{TIS_BAi_HighSpeedRailway,CST_CXWang_HighSpeed} or on vehicles with a speed of up to 300 km/h \cite{CM_JChoi_VehicularCommunication,JSAC_LDai_HighSpeed}. Currently, the dominant modulation technique for 4G and the emerging 5G is orthogonal frequency division multiplexing (OFDM). For the high-mobility scenarios, OFDM may experience significant inter-carrier interference (ICI) due to the Doppler spread of time-variant channels (which are also referred to as doubly selective or doubly dispersive channels). ICI will severely degrade the performance of OFDM systems when the traditional transceivers are used \cite{CM_HSair_transmission1995}.
	
	To cope with ICI, some modifications of the traditional OFDM were proposed at the cost of more complicated transceiver design. Linear equalization \cite{TCOM_WJeon_anequalization1999,TCOM_XCai_Bounding2003,TSP_PSchniter_low2004,TSP_SDas_maxsinr2007} and non-linear equalization \cite{TCOM_YChoi_Onchannel2001,TVT_AMolisch_iterative2007,TSP_KFang_low2008} were proposed to eliminate the ICI at the receiver. Some transmitter processing methods to mitigate ICI were proposed including polynomial cancellation coding \cite{TCOM_YZhao_intercarrier2001,TIT_KASeaton_Polynomial2000} and pulse shaping \cite{JSAC_WKozek_nonothogonal1998,TIT_KLiu_STF}. Using both transmitter and receiver processing, a channel-independent block spreading based multiple access scheme was proposed for the multi-user scenarios, where both ICI and multiuser interference can be eliminated \cite{TIT_GLeus_BlockSpreading}.
	
	Instead of trying to eliminate ICI, there are some modulation schemes proposed for time-variant channels to enhance the system performance by using the transmit diversity. A frequency-oversampling technique for zero-padded OFDM system was proposed in \cite{JOE_ZWang_D-OFDM}, where frequency diversity can be achieved through transmit signal design. Vector OFDM \cite{TCOM_XGXia_VOFDM} transmits multiple groups of linearly precoded symbols over the channel subcarriers to provide frequency diversity. A Doppler-resilient orthogonal signal division multiplexing technique was proposed. That multiplexes several data vectors and a pilot vector into a data stream to fully exploit the frequency-time diversity in the time-variant channels \cite{JOE_TEbihara_OSDM_Doppler,JOE_TEbihara_OSDM}.
	
	Orthogonal time frequency space (OTFS) is an alternative to OFDM to tackle the time-variant channels \cite{WCNC_RHadani_OTFS,IMS_RHadani_OTFS,WCL_AFarhang_LowComplexityOTFS}. Leveraging the basis expansion model (BEM) for the channel \cite{Proc_Giannakis_BEM,TCOM_PBello_BEM}, OTFS converts the time-variant channels into the \textit{time-independent} channels in the delay-Doppler domain. Accordingly, the information bearing data is multiplexed into the roughly constant channels in the delay-Doppler domain. OTFS is different from previous work in that it multiplexes data in the delay-Doppler domain.
	
	Like OFDM multi-antenna systems, OTFS with massive multiple-input multiple-output (MIMO) can further increase the spectrum efficiency. Such benefits require that the channel state information (CSI) is known at the transmitter to design the transmit beamforming vectors \cite{arxiv_hadani2018otfs1,arxiv_hadani2018otfs,arxiv_li2017simple}. When OTFS massive MIMO systems are operated with frequency division duplex (FDD) mode, downlink channel estimation is necessary due to the lack of channel reciprocity. With a large number of antennas equipped at the base station (BS) in OTFS massive MIMO systems, downlink channel estimation is challenging.
	
	The time-variant channel estimation schemes for massive MIMO systems have been proposed in \cite{TCOM_PCheng_BEM_CS,TVT_HXie_Unified,Proc_Bajwa_Double_Selective_CS}. In \cite{TCOM_PCheng_BEM_CS}, the time-variant MIMO channels are modeled by several jointly sparse time-independent coefficients based on the BEM. These jointly sparse coefficients can be estimated through a distributed compressive sensing algorithm with high accuracy. In \cite{TVT_HXie_Unified}, a spatial-domain BEM was developed to further reduce the effective dimensions of massive MIMO time-variant channels, such that the downlink training overhead can be reduced. Moreover, a general framework of compressed channel sensing was provided in \cite{Proc_Bajwa_Double_Selective_CS}. Based on the sparse multipath
	structure of massive MIMO time-variant channels, compressed channel sensing can achieve a target estimation error using much less overhead. The aforementioned channel estimation techniques \cite{TCOM_PCheng_BEM_CS,TVT_HXie_Unified,Proc_Bajwa_Double_Selective_CS} were proposed for OFDM massive MIMO systems. They are not directly applicable for OTFS massive MIMO systems. This is because that the information bearing data is multiplexed in the delay-Doppler domain in OTFS systems, not the frequency-time domain as in OFDM systems.
	
	For OTFS systems, an impulse based channel estimation technique was proposed for the OTFS single-input single-output (SISO) architecture in \cite{arxiv_Monk_OTFS}. The BS transmits an impulse in the delay-Doppler domain as the training pilots. The received signals in the delay-Doppler domain
	can be regarded as a two-dimensional periodic convolution of the transmit impulse with the delay-Doppler channel \cite{arxiv_Monk_OTFS}. The delay-Doppler channel can then be estimated from the received signal. An alternative method using PN sequences as the training pilots in the delay-Doppler
	domain was proposed for OTFS SISO systems \cite{arxiv_murali2018otfs}. In that method, channel estimation is done in the discrete domain, where three quantities of interest, namely, delay shift, Doppler shift, and fade coefficient are estimated. Then the delay-Doppler channel can be calculated accordingly.
	The impulse-based scheme is extended to OTFS MIMO systems by transmitting several impulses with proper guard between two adjacent impulses to distinguish different BS antennas \cite{arxiv_ramachandran2018mimo}. The existing channel estimation techniques can not be directly extended to OTFS massive MIMO
	since a large number of antennas are required to be distinguished by transmitting orthogonal pilots, which will lead to high pilot overhead.

	To solve this problem, we propose a 3D structured orthogonal matching pursuit (3D-SOMP) algorithm based downlink channel estimation technique for OTFS massive MIMO systems, which can achieve accurate CSI with low pilot overhead. The specific contributions are summarized as follows.
	\begin{itemize}
		\item
		We present the discrete-time formulation of OTFS systems and demonstrate that the OTFS massive MIMO channel exhibits a delay-Doppler-angle 3D structured sparsity. Since the number of dominant propagation paths is limited, the 3D channel is sparse along the delay dimension. As the Doppler frequency of a path is usually much smaller than the system bandwidth, the 3D channel is block-sparse along the Doppler dimension. The only one non-zero block is concentrated around zero, but the length of the non-zero block is unknown. Since the angle-of-departure (AoD) spread of a path at the BS is usually small, the 3D channel is burst-sparse along the angle dimension \cite{TWC_ALiu_BurstSparsity}. The lengths of non-zero bursts can be regarded as constant, but the start position of each non-zero burst is unknown.
		\item
		Based on the 3D structured sparse channel, we formulate the downlink channel estimation problem in OTFS massive MIMO systems as a sparse signal recovery problem. The estimator makes use of the training pilots that are transmitted in the delay-Doppler domain. We propose that pilots of different antennas are independent complex Gaussian random sequences, which overlap to reduce the overall pilot overhead. By inserting guard intervals between pilots and data, the received pilots can be expressed as a phase compensated two-dimensional periodic convolution of the transmit pilots with the delay-Doppler channel. Decomposing
		the channel based on its structure, we formulate the downlink channel estimation problem as a sparse signal recovery problem.
		\item
		We propose a 3D-SOMP algorithm to solve the formulated channel estimation problem. The main idea is summarized as follows. The 3D support of each path is estimated in an one-by-one fashion. For each path, the user first estimates the delay-dimension support. Then, by using the block-sparse property of channels along the Doppler dimension, the user estimates the size of the unique non-zero block to obtain the Doppler-dimension support. Finally, the user transforms the burst-sparsity of channels along the angle dimension into the traditional block-sparsity through a lifting transformation following \cite{TWC_ALiu_BurstSparsity}, so that the angle-dimension support can be estimated accordingly. In this way, the whole 3D channel can be estimated after several iterations by removing the contribution of previous paths in each iteration.
	\end{itemize}
	
	The rest of the paper is organized as follows. In Section II, we present the system model. In Section III, we review the channel estimation in OTFS SISO systems. Then, we propose a 3D-SOMP based channel estimation technique for OTFS massive MIMO systems in Section IV. Simulation results are given in Section V. Our conclusions are finally drawn in Section VI.
	
	\textit{Notation}: Boldface capital letters stand for matrices and lower-case letters stand for column vectors. The transpose, conjugate, conjugate transpose, and inverse of a matrix are denoted by $(\cdot)^{\rm T}$, $(\cdot)^*$, $(\cdot)^{\rm H}$ and $(\cdot)^{-1}$, respectively. $\odot$ is the Hadamard product operator. $\|\mathbf{s}\|$ is the $\ell_2$-norm of the vector $\mathbf{s}$. $\mathbf{\Psi}^{\dagger}=(\mathbf{\Psi}^{\rm H}\mathbf{\Psi})^{-1}\mathbf{\Psi}^{\rm H}$ is the Moore-Penrose pseudo-inverse of $\mathbf{\Psi}$. Finally, $\mathbf{I}_N$ denotes the identity matrix of size $N\times N$.
	
	\section{System Model}\label{S2}
	In this section, we review OTFS for SISO systems including a discrete-time formulation of OTFS modulation and OTFS demodulation. Then, we describe an extension of OTFS into massive MIMO systems.
	
\begin{figure*}[t] 
		\vspace{-0mm}
		\center{\includegraphics[width=1.8\columnwidth]{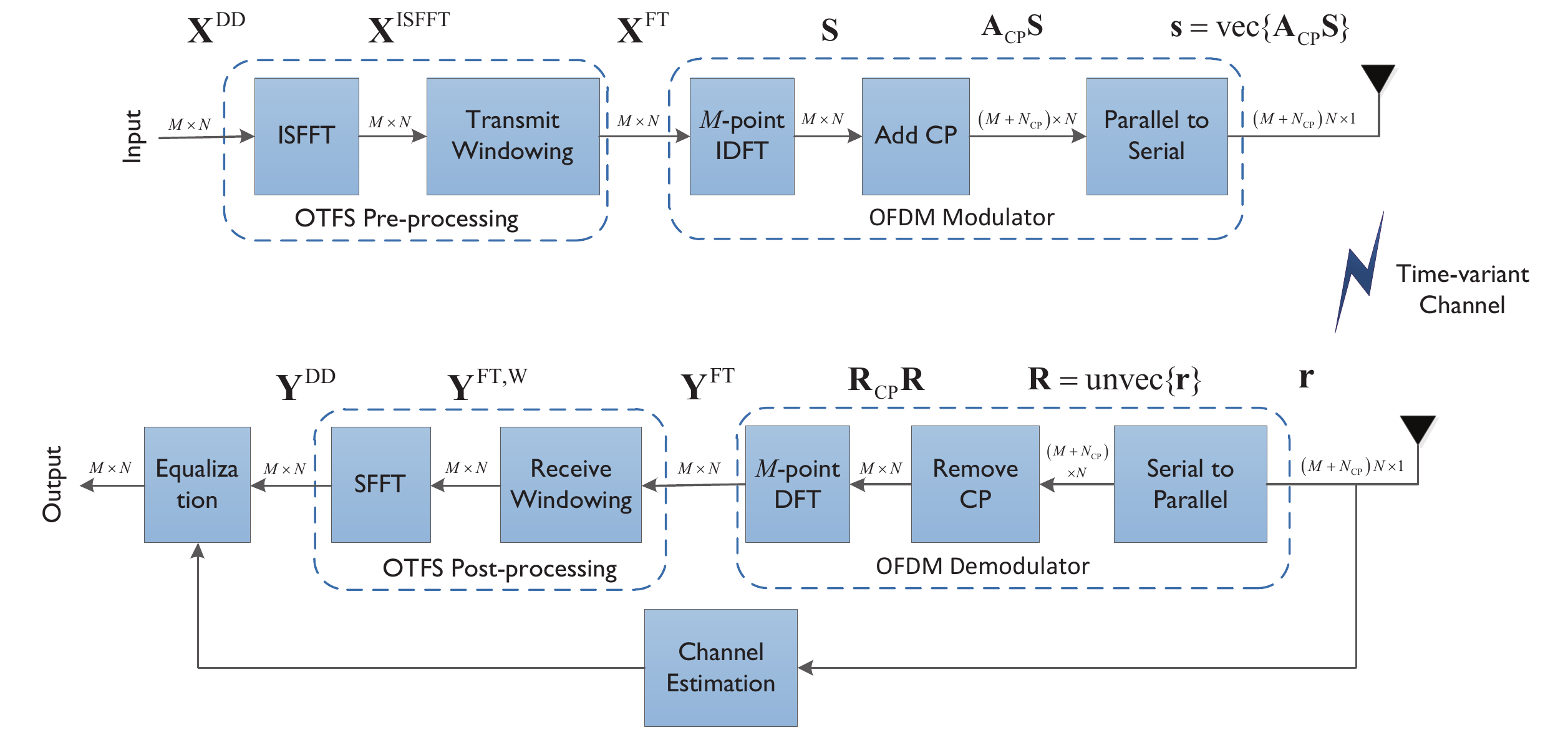}}
		\vspace{-0mm}
		\caption{OTFS SISO architecture. OTFS modulation is composed of a pre-processing block before a traditional OFDM modulator at the transmitter. OTFS demodulation is composed of a post-processing block after a traditional OFDM demodulator at the receiver.}
		\label{Fig_OTFS_Architecture}
	\end{figure*}

	Fig. \ref{Fig_OTFS_Architecture} shows the OTFS SISO architecture as commonly assumed in \cite{WCNC_RHadani_OTFS,IMS_RHadani_OTFS,WCL_AFarhang_LowComplexityOTFS}. OTFS is a modulation/demodulation technique. It can be realized by adding a pre-processing block before a traditional modulator in the frequency-time domain such as OFDM modulator at the transmitter, and a corresponding post-processing block after a traditional demodulator in the frequency-time domain such as OFDM demodulator at the receiver. Through the pre-processing and post-processing blocks, time-variant channels are converted into the time-independent channels in the the delay-Doppler domain. Therefore, the information bearing data can be multiplexed in the roughly constant delay-Doppler channel. At the same time, the transmit data in OTFS systems can take advantage of full diversity in the frequency-time channels. In this way, OTFS improves system performance over OFDM in high-mobility scenarios \cite{WCNC_RHadani_OTFS,IMS_RHadani_OTFS,WCL_AFarhang_LowComplexityOTFS}.
	
	\subsection{OTFS SISO Modulation}\label{S2.1}
	In this section, we describe the modulation at the transmitter. A quadrature amplitude modulated (QAM) data sequence of length $MN$ is first rearranged into a 2D data block. This is called a 2D OTFS frame in the delay-Doppler domain $\mathbf{X}^{\rm DD}\in\mathbb{C}^{M \times N}$, where $M$ and $N$ are the numbers of resource units along the delay dimension and Doppler dimension. OTFS modulation at the transmitter is composed of a pre-processing block and a traditional frequency-time modulator such as OFDM or filter bank multicarrier (FBMC). The pre-processing block maps the 2D data block $\mathbf{X}^{\rm DD}$ in the delay-Doppler domain to a 2D block $\mathbf{X}^{\rm FT}\in\mathbb{C}^{M \times N}$ in the frequency-time domain. It is realized by using an inverse symplectic finite Fourier transform (ISFFT) and a transmit windowing function. The ISFFT of $\mathbf{X}^{\rm DD}$ is \cite{WCNC_RHadani_OTFS}
	\begin{align}\label{eq_XISFFT} 
	\mathbf{X}^{\rm ISFFT}=\mathbf{F}_{\rm M}\mathbf{X}^{\rm DD}\mathbf{F}_{\rm N}^{\rm H},
	\end{align}
	where $\mathbf{F}_{\rm M}\in\mathbb{C}^{M \times M}$ and $\mathbf{F}_{\rm N}\in\mathbb{C}^{N \times N}$ are discrete Fourier transform (DFT) matrices. 
	A transmit windowing matrix $\mathbf{W}^{\rm tx}\in\mathbb{C}^{M \times N}$ multiplies $\mathbf{X}^{\rm ISFFT}$ element-wise to produce the 2D block in the frequency-time domain $\mathbf{X}^{\rm FT}$ as
	\begin{align}\label{eq_Xnm} 
	\mathbf{X}^{\rm FT}=\mathbf{X}^{\rm ISFFT}\odot\mathbf{W}^{\rm tx}.
	\end{align}
	There are several uses of the windowing matrix. For example, the windowing matrix can be designed to randomize the phases of the transmitted symbols to eliminate the inter-cell interference \cite{arxiv_Monk_OTFS}. In this paper, we assume a trivial window at the transmitter for simple expression, i.e., $\mathbf{W}^{\rm tx}$ is a matrix of all ones.
	
	Then, the 2D block $\mathbf{X}^{\rm FT}$ in the frequency-time domain is transformed to the 1D transmit signal $\mathbf{s}$ through a traditional frequency-time modulator such as OFDM or FBMC. Assuming an OFDM modulator, the $M$-point inverse DFT (IDFT) is applied on each column of $\mathbf{X}^{\rm FT}$ to obtain the 2D transmit signal block $\mathbf{S}\in\mathbb{C}^{M\times N}$, i.e.,
	\begin{align}\label{eq_S} 
	\mathbf{S}=\mathbf{F}_{\rm M}^{\rm H}\mathbf{X}^{\rm FT},
	\end{align}
	where $\mathbf{S}=\left[ \mathbf{s}_1,\mathbf{s}_2,\cdots,\mathbf{s}_N \right] $. Each column vector $\mathbf{s}_i\in\mathbb{C}^{M\times 1}$ of $\mathbf{S}$ can be regarded as an OFDM symbol. Note that $N$ OFDM symbols $\{\mathbf{s}_i\}_{i=1}^{N}$ occupy the bandwidth $M\Delta f$ and have the duration $NT$, where $\Delta f$ and $T$ are the subcarrier spacing and symbol duration. By combing (\ref{eq_XISFFT})-(\ref{eq_S}),
	\begin{align}\label{eq_S1} 
	\mathbf{S}=\mathbf{X}^{\rm DD}\mathbf{F}_{\rm N}^{\rm H}.
	\end{align}
	To avoid inter-symbol interference between blocks, the OFDM modulator usually adds cyclic prefix (CP) for each OFDM symbol $\mathbf{s}_i$ via a CP addition matrix $\mathbf{A}_{\rm CP}\in\mathbb{C}^{(M+N_{\rm CP})\times M}$\cite{WCL_AFarhang_LowComplexityOTFS} with $N_{\rm CP}$ being the length of CP. By reading the 2D transmit signal block $\mathbf{S}$ column-wise, the 1D transmit signal $\mathbf{s}\in\mathbb{C}^{(M+N_{\rm CP})N\times 1}$ is
	\begin{align}\label{eq_s} 
	\mathbf{s}={\rm vec}\{\mathbf{A}_{\rm CP}\mathbf{S}\}.
	\end{align}
	
	\subsection{OTFS SISO Demodulation}\label{S2.2}
	In this section, we describe demodulation at the receiver. The $\kappa$-th element of the received signal $\mathbf{r}\in\mathbb{C}^{(M+N_{\rm CP})N\times 1}$ after the time-variant channel $h_{\kappa,\ell}$ with length $L+1$ is expressed as 
	\begin{align}\label{eq_rt} 
	r_\kappa = \sum_{\ell=0}^{L} h_{\kappa,\ell}s_{\kappa-\ell} +v_\kappa,
	\end{align}
	where $v_\kappa$ is the additive noise at the receiver. The OTFS demodulation at the receiver consists of a traditional frequency-time demodulator such as the OFDM or FBMC demodulator and a post-processing block as shown in Fig. \ref{Fig_OTFS_Architecture}. The frequency-time demodulator transforms the received
	signal $\mathbf{r}$ to a 2D block in the frequency-time domain $\mathbf{Y}^{\rm FT}\in\mathbb{C}^{M \times N}$. Specifically, assuming an OFDM demodulator, the received signal $\mathbf{r}$ is first rearranged as a matrix $\mathbf{R}$ of size $(M+N_{\rm CP})\times N$, i.e.,
	\begin{align}\label{eq_R} 
	\mathbf{R}={\rm unvec}\{\mathbf{r}\},
	\end{align}
	where each column vector of $\mathbf{R}$ can be regarded as a received OFDM symbol including CP. Then, the OFDM demodulator removes the CP by multiplying $\mathbf{R}$ with a CP removal matrix $\mathbf{R}_{\rm CP}\in\mathbb{C}^{M\times (M+N_{\rm CP})}$ \cite{WCL_AFarhang_LowComplexityOTFS} to obtain the OFDM symbols $\mathbf{R}_{\rm CP}\mathbf{R}$ without CPs. Applying the $M$-point DFT on each OFDM symbol without CP (i.e., each column vector of $\mathbf{R}_{\rm CP}\mathbf{R}$), we obtain the received 2D block $\mathbf{Y}^{\rm FT}$ in the frequency-time domain as
	\begin{align}\label{eq_R} 
	\mathbf{Y}^{\rm FT}=\mathbf{F}_{\rm M}\mathbf{R}_{\rm CP}\mathbf{R}.
	\end{align}
	
	In the post-processing block, $\mathbf{Y}^{\rm FT}$ is transformed to the 2D data block $\mathbf{Y}^{\rm DD}\in\mathbb{C}^{M\times N}$ in the delay-Doppler domain. It is realized by a receive windowing matrix $\mathbf{W}^{\rm rx}\in\mathbb{C}^{M \times N}$ and the SFFT. The receive windowing matrix $\mathbf{W}^{\rm rx}$ multiplies $\mathbf{Y}^{\rm FT}$ element-wise, i.e.,
	\begin{align}\label{eq_YWnm} 
	\mathbf{Y}^{\rm FT,W}=\mathbf{Y}^{\rm FT}\odot\mathbf{W}^{\rm rx}.
	\end{align}
	Then, the SFFT is applied for $\mathbf{Y}^{\rm FT,W}$ to obtain the 2D data block $\mathbf{Y}^{\rm DD}$ in the delay-Doppler domain as
	\begin{align}\label{eq_YDD} 
	\mathbf{Y}^{\rm DD}=\mathbf{F}_{\rm M}^{\rm H}\mathbf{Y}^{\rm FT,W}\mathbf{F}_{\rm N}.
	\end{align}
	Like the transmitter, we consider a trivial window at the receiver for simple expression, i.e., $\mathbf{W}^{\rm rx}$ is a matrix of all ones \cite{arxiv_Monk_OTFS}. By combing (\ref{eq_R})-(\ref{eq_YDD}), we can obtain
	\begin{align}\label{eq_YDD1} 
	\mathbf{Y}^{\rm DD}=\mathbf{R}_{\rm CP}\mathbf{R}\mathbf{F}_{\rm N}.
	\end{align}
	The received 2D data block $\mathbf{Y}^{\rm DD}$ in the delay-Doppler domain is given by the phase compensated two-dimensional periodic convolution of the transmit 2D data block $\mathbf{X}^{\rm DD}$ in the delay-Doppler domain with the delay-Doppler channel impulse response (CIR) $\mathbf{H}^{\rm DD}\in\mathbb{C}^{M\times N}$ as shown in the
	following Lemma 1.
	
	\textbf{Lemma 1}:
	We denote the $(\ell+1,k+1+N/2)$-th element of $\mathbf{Y}^{\rm DD}$ and $\mathbf{X}^{\rm DD}$ as $Y^{\rm DD}_{\ell,k}$ and $X^{\rm DD}_{\ell,k}$, where $\ell=0,1,\cdots,M-1$ and $k=-N/2,\cdots,0,\cdots,N/2-1$. Then $Y^{\rm DD}_{\ell,k}$ can be expressed as 
	\begin{align}\label{eq_YDDlk2} 
	Y^{\rm DD}_{\ell,k}\overset{N\rightarrow\infty}{=}\sum_{\ell'=0}^{M-1}\sum_{k'=-N/2}^{N/2-1} &X^{\rm DD}_{\ell',k'}H^{\rm DD}_{\ell-\ell',k-k'}e^{j2\pi\frac{\ell \left( k-k'\right)}{N(M+N_{\rm CP})}}\\\nonumber &+V^{\rm DD}_{\ell,k},
	\end{align}
	where $V^{\rm DD}_{\ell,k}$ is the additive noise in the delay-Doppler domain. $H^{\rm DD}_{\ell,k}$ is the $(\ell+1,k+1+N/2)$-th element of the delay-Doppler CIR $\mathbf{H}^{\rm DD}$ and 
	\begin{align}\label{eq_HDD} 
	H^{\rm DD}_{\ell,k}=\sum_{i=1}^{N}h_{(i-1)(M+N_{\rm CP})+1,(\ell)_M}e^{-j2\pi (i-1)\frac{k}{N}},
	\end{align}
	where $(\ell)_M$ is the remainder after division of $\ell$ by $M$. Note that $H^{\rm DD}_{\ell,k}=H^{\rm DD}_{\ell+M,k+N}$, thus (\ref{eq_YDDlk2}) can be regarded as periodic convolution.
	\begin{IEEEproof}
		See Appendix I.
	\end{IEEEproof}
	
	We observe from (\ref{eq_YDDlk2}) that the transmit data $X^{\rm DD}_{\ell',k'}$ in the delay-Doppler domain experiences roughly constant channel $H^{\rm DD}_{\ell,k}$ in the delay-Doppler domain, since the delay-Doppler CIR $\mathbf{H}^{\rm DD}$ is time-independent ($\mathbf{H}^{\rm DD}$ does not vary with the variable $\kappa$ of $h_{\kappa, \ell}$). Moreover, since each transmit data $X^{\rm DD}_{\ell',k'}$ in the delay-Doppler domain is expanded onto the whole frequency-time domain as shown in (\ref{eq_XISFFT}) and (\ref{eq_Xnm}), it can exploit the full diversity of the frequency-time channel. As a result, OTFS has improved performance over the traditional OFDM especially in high-mobility scenarios\cite{WCNC_RHadani_OTFS,IMS_RHadani_OTFS,WCL_AFarhang_LowComplexityOTFS}.
	
	Equalization is required to eliminate the inter-symbol interference, since each transmit data $X^{\rm DD}_{\ell',k'}$ in (\ref{eq_YDDlk2}) experiences not only the delay-Doppler channel $X^{\rm DD}_{\ell',k'}H^{\rm DD}_{0,0}$ but also the inter-symbol interference $X^{\rm DD}_{\ell',k'}H^{\rm DD}_{\ell-\ell',k-k'},\forall \ell'\neq \ell,k'\neq k$. To eliminate such inter-symbol interference through equalization, the delay-Doppler CIR $\mathbf{H}^{\rm DD}$ is required, which is obtained through downlink channel estimation.
	
	\subsection{OTFS Massive MIMO}
	\begin{figure*}[ht!] 
		\vspace{-0mm}
		\center{\includegraphics[width=1.8\columnwidth]{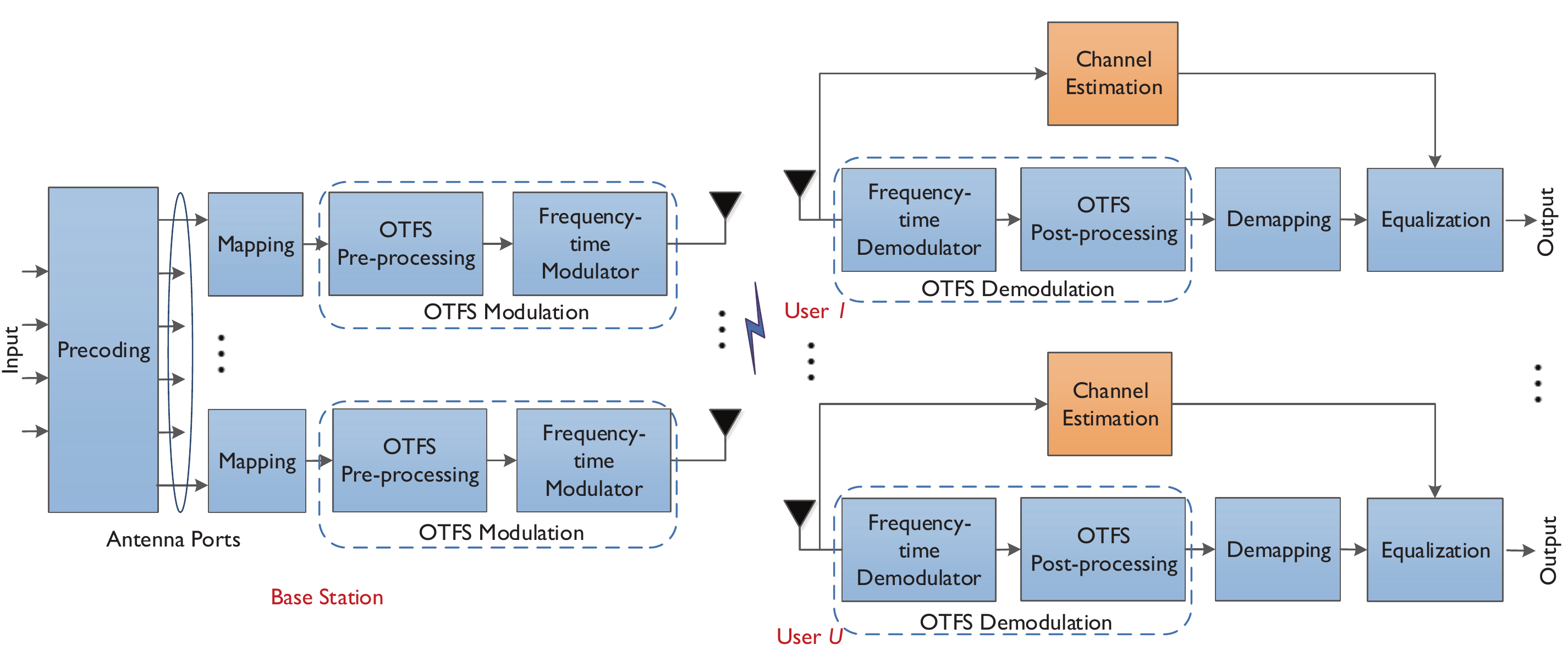}}
		\vspace{-0mm}
		\caption{OTFS massive MIMO architecture. Multi-user MIMO is used to increase the spectrum efficiency. Downlink precoding is performed based on the downlink CSI\cite{arxiv_hadani2018otfs}, which is obtained through downlink channel estimation and uplink channel feedback in FDD systems.}
		\label{Fig_OTFS_MassiveMIMO}
		\vspace{-0mm}
	\end{figure*}
	
	We explain how OTFS work in massive MIMO systems to further increase the spectrum
	efficiency by using multi-user MIMO in this section. Fig. \ref{Fig_OTFS_MassiveMIMO} shows the OTFS massive MIMO architecture. The BS is equipped with $N_{\rm t}$ antennas to simultaneously serve $U$ single-antenna users. Downlink precoding is performed to eliminate the inter-user interference. For example, the zero-forcing Tomlinson-Harashima precoding is adopted in \cite{arxiv_hadani2018otfs}. To perform downlink precoding, downlink CSI is required, which is obtained from uplink channel feedback in FDD systems. After precoding, the transmit data block $\mathbf{X}^{\rm DD}$ in the delay-Doppler domain will be modulated through the OTFS modulation and transmitted at $N_{\rm t}$ antennas. At the user side, the received signal is first demodulated through the OTFS demodulation to obtain the received data block $\mathbf{Y}^{\rm DD}$ in the delay-Doppler domain. To cancel the inter-symbol interference, equalization is performed based on the downlink CSI. Next, we will focus on the downlink channel estimation in OTFS SISO/massive MIMO systems.
	
	\section{Channel Estimation in OTFS SISO Systems}\label{S3}
	The goal of channel estimation is to obtain the delay-Doppler CIR $\mathbf{H}^{\rm DD}$ from the received delay-Doppler data block $\mathbf{Y}^{\rm DD}$ in (\ref{eq_YDDlk2}). One intuitive method to estimate $\mathbf{H}^{\rm DD}$ is to transmit an impulse in the delay-Doppler domain as the training pilots\cite{arxiv_Monk_OTFS}. The transmit impulse is expressed as 
	\begin{align}\label{eq_Xklp}
	X_{\ell,k }^{{\rm DD}} = \left\{ \begin{array}{l}
	1,\quad \ell  = 0, k = 0,\\
	0,\quad \ell  \neq 0, k \neq 0.
	\end{array} \right. 
	\end{align}
	Based on (\ref{eq_YDDlk2}), the received signal in the delay-Doppler domain can be expressed as
	\begin{align}\label{eq_Ykl4} 
	Y^{\rm DD}_{\ell,k}=H^{\rm DD}_{\ell,k}e^{j2\pi\frac{\ell k}{N(M+N_{\rm CP})}}+V^{\rm DD}_{\ell,k},
	\end{align}
	The delay-Doppler CIR $H^{\rm DD}_{\ell,k}$ can be estimated from $Y^{\rm DD}_{\ell,k}$ in (\ref{eq_Ykl4}) through the least square (LS) estimator or minimum mean square error (MMSE) estimator \cite{arxiv_ramachandran2018mimo}. Note that only the non-zero part of $H^{\rm DD}_{\ell,k}$ need to be estimated due to its finite support, which will be explained later.
	
	This impulse based channel estimation technique, however, is not applicable to massive MIMO systems due to the huge required pilot overhead. In OTFS massive MIMO systems, to distinguish the delay-Doppler channels associated with $N_{\rm t}$ BS antennas at the user side, $N_{\rm t}$ impulses are required to be transmitted. We assume that the delay-Doppler CIRs $H^{\rm DD}_{\ell,k}$ of $N_{\rm t}$ antennas have finite support $[0:M_{\rm max}-1]$ along the delay dimension and $\left[-\frac{N_{\rm max}}{2}:\frac{N_{\rm max}}{2}-1\right]$ along the Doppler dimension \cite{WCNC_RHadani_OTFS,arxiv_Monk_OTFS,arxiv_hadani2018otfs}. To avoid the interference among multiple antennas, guard intervals between two adjacent impulses should not be smaller than $N_{\rm max}$ along the Doppler dimension and no smaller than $M_{\rm max}$ along the delay dimension \cite{arxiv_Monk_OTFS}. As a result, the length of pilots to transmit $N_{\rm t}$ impulses in OTFS massive MIMO systems should be $\propto N_{\rm t}N_{\rm max}M_{\rm max}$. With a large number of BS antennas, the pilot overhead will be overwhelming. To solve this problem, we propose a 3D-SOMP algorithm based channel estimation technique, which can obtain the accurate CSI with considerably reduced pilot overhead.

	\section{Proposed 3D-SOMP Based Channel Estimation in OTFS Massive MIMO Systems}\label{S4}
	In this section, we first demonstrate the 3D structured sparsity of channels in OTFS massive MIMO systems. Then, we formulate the downlink channel estimation problem as a sparse signal recovery problem. To solve this problem, we propose a 3D-SOMP algorithm. Finally, we analyze the required pilot overhead for the proposed 3D-SOMP based channel estimation technique.
	
	\subsection{3D Structured Sparsity of Delay-Doppler-angle Channel}\label{S4.1}
	We consider an OTFS massive MIMO system with $N_{\rm t}$ antennas at the BS and $U$ single-antenna users. Downlink channel estimation is the same for $U$ users. Therefore, we focus on a certain user and omit the subscript for the user without loss of generality. We consider the downlink time-variant channel consisting of $N_{\rm p}$ dominant propagation paths. Each dominant path is composed of $N_{\rm s}$ subpaths. The $s_i$-th subpath in the $i$-th dominant path has a complex path gain $\alpha_{s_i}$ and Doppler frequency $\nu_{s_i}$. The delays of all subpaths in the $i$-th dominant path can be regarded as the same $\tau_i$\cite{3GPPTR_SCM}. We denote the physical AoD of the $s_i$-th subpath as $\theta_{s_i}$. When a typical uniform linear array (ULA) of antennas is considered, the spatial angle associated with $\theta_{s_i}$ is defined as $\psi_{s_i}=\frac{d}{\lambda}\sin\theta_{s_i}$ \cite{JTSP_HRobert_OverviewMmwave}, where $d$ is the antenna spacing and $\lambda$ is the wavelength of the carrier frequency. Typically, $d = \lambda/2$ and $\theta_{s_i}\in[-\pi/2,\pi/2)$, thus $\psi_{s_i}\in[-1/2,1/2)$. The time-variant channel associated with the $(p+1)$-th antenna ($p=0,1,\cdots, N_{\rm t}-1$) can be expressed as\cite{AC_FHlawatsch_WirelessRapidly}
	\begin{align}\label{eq_hkappalp} 
	h_{\kappa,\ell,p}=\sum_{i=1}^{N_{\rm p}}\sum_{s_i=1}^{N_{\rm s}} \alpha_{s_i}e^{j2\pi\nu_{s_i}\kappa T_{\rm s}}{\rm p_{rc}}(\ell T_{\rm s}-\tau_i)e^{-j2\pi p\psi_{s_i}},
	\end{align}
	where ${\rm p_{rc}}(\tau)$ is the band-limited pulse shaping filter response evaluated at $\tau$ and $T_{\rm s}=\frac{1}{M\Delta f}$ is the system sampling interval.
	Based on (\ref{eq_HDD}), we express the delay-Doppler CIR of the $(p+1)$-th antenna (which is referred to as delay-Doppler-space CIR $H^{\rm DDS}_{\ell,k,p}$ in OTFS massive MIMO systems, where $\ell$, $k$ and $p$ correspond to the delay, Doppler and spatial index) as follows 
	\begin{align}\label{eq_DDS} 
	&H^{\rm DDS}_{\ell,k,p}=\sum_{n=1}^{N}h_{(n-1)(M+N_{\rm CP})+1,(\ell)_M,p}e^{-j2\pi (n-1)\frac{k}{N}} \\\nonumber
	&=\sum_{i=1}^{N_{\rm p}}\sum_{s_i=1}^{N_{\rm s}} \beta_{s_i}\Upsilon_N\left( \nu_{s_i}NT-k\right) {\rm p_{rc}}\left( (\ell)_M T_{\rm s}-\tau_i\right) e^{-j2\pi p\psi_{s_i}},
	\end{align}
	where $\beta_{s_i}=\alpha_{s_i}e^{j2\pi\nu_{s_i}T_{\rm s}}$, $\Upsilon_N(x)\triangleq\sum_{n=1}^{N}e^{j2\pi\frac{x}{N}(n-1)}=\frac{\sin(\pi x)}{\sin(\pi \frac{x}{N})}e^{j\pi \frac{x(N-1)}{N}}$ and $T=(M+N_{\rm CP})T_{\rm s}$.
	
	To investigate the 3D structured sparsity of channels in OTFS massive MIMO systems, we define the delay-Doppler-angle channel $H^{\rm DDA}_{\ell,k,r}$ by applying inverse DFT for $H^{\rm DDS}_{\ell,k,p}$ along the space-dimension $p$ as
	\begin{align}\label{eq_hklr1} 
	H^{\rm DDA}_{\ell,k,r}\overset{\Delta}{=} \sum_{p=0}^{N_{\rm t}-1}H^{\rm DDS}_{\ell,k,p}e^{j2\pi \frac{rp}{N_{\rm t}}}
	\end{align}
	where $r=-\frac{N_{\rm t}}{2},\cdots,0,\cdots,\frac{N_{\rm t}}{2}-1$ is the angle index. Then, by substituting (\ref{eq_DDS}) into (\ref{eq_hklr1}), we can express the delay-Doppler-angle channel $H^{\rm DDA}_{\ell,k,r}$ as
	\begin{align}\label{eq_hklr2} 
	H^{\rm DDA}_{\ell,k,r}= \sum_{i=1}^{N_{\rm p}}\sum_{s_i=1}^{N_{\rm s}}& \beta_{s_i} \Upsilon_N\left( \nu_{s_i}NT-k\right)\\\nonumber&\times{\rm p_{rc}}\left( (\ell)_M T_{\rm s}-\tau_i\right) \Upsilon_{N_{\rm t}}(r-\psi_{s_i}N_{\rm t}).
	\end{align}
	We arrange $H^{\rm DDA}_{\ell,k,r}$ into a 3D tensor $\mathcal{H}\in\mathbb{C}^{M\times N\times N_{\rm t}}$, where $H^{\rm DDA}_{\ell,k,r}$ is the $(\ell+1,k+N/2+1,r+N_{\rm t}/2+1)$-th element of $\mathcal{H}$ ($\ell=0,1,\cdots,M-1$, $k=-N/2,\cdots,0,\cdots,N/2-1$, and $r=-N_{\rm t}/2,\cdots,0,\cdots,N_{\rm t}/2$).
	
	\begin{figure}[t] 
		\vspace{-0mm}
		\center{\includegraphics[width=0.9\columnwidth]{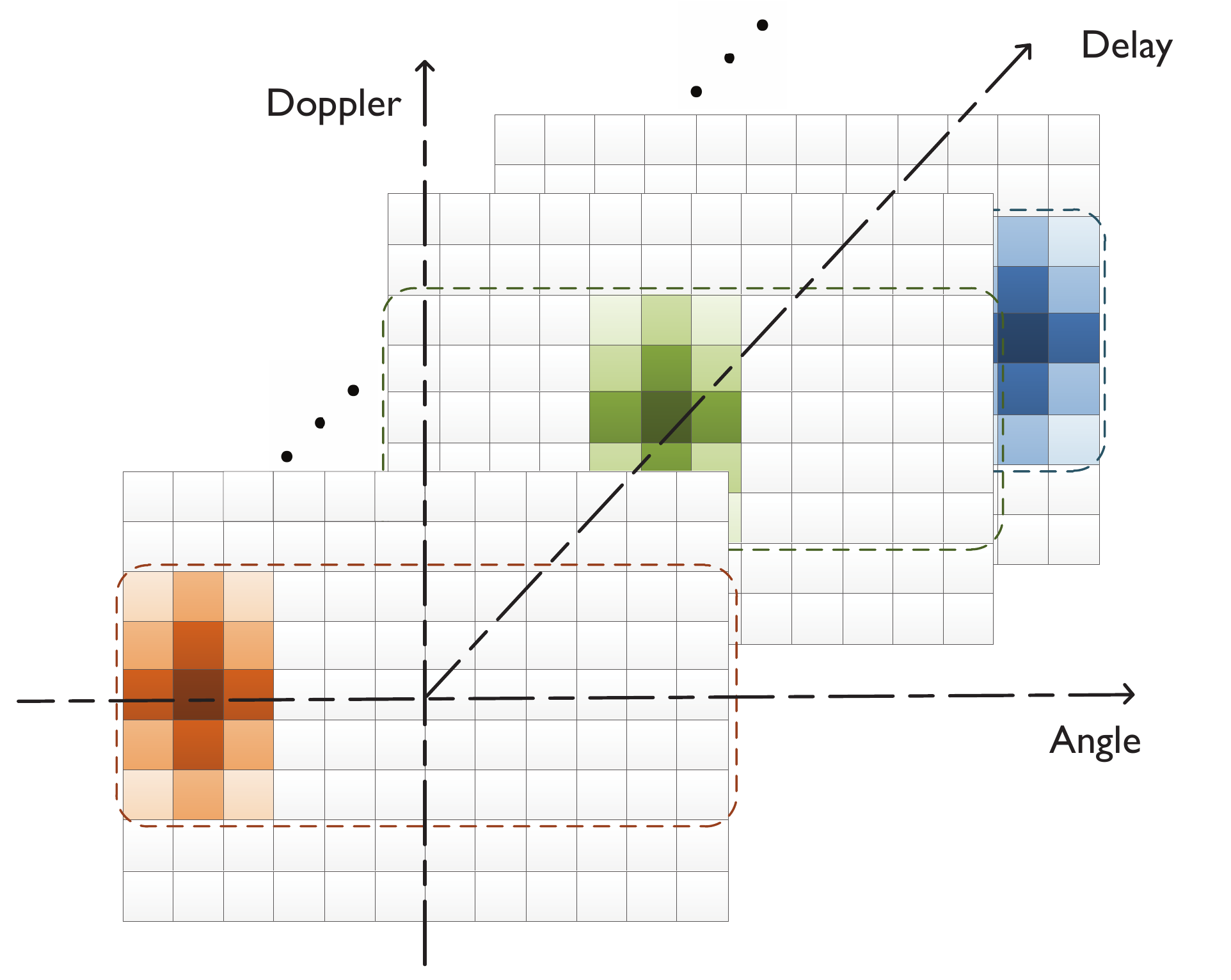}}
		\vspace{-0mm}
		\caption{Delay-Doppler-angle 3D channel, which is sparse along the delay dimension, block-sparse along the Doppler dimension, and burst-sparse along the angle dimension.}
		\label{Fig_DDA_Channel}
		\vspace{-0mm}
	\end{figure}
	
	The function $\Upsilon_N(x)$ has the following characteristic: $|\Upsilon_N(x)|\approx 0$ when $|x|\gg 1$\cite{TWC_XGao_BeamspaceChannelEstimation}. Therefore, $H^{\rm DDA}_{\ell,k,r}$ has dominant elements only if $k\approx\nu_{s_i}NT$, $\ell\approx\tau_iM\Delta f$, and $r\approx\psi_{s_i}N_{\rm t}$. As shown in Fig. \ref{Fig_DDA_Channel}, since the number of dominant paths is small, e.g., $N_{\rm p}=6$\cite{3GPPTR_SCM} (the path delays of $N_s$ subpaths of a dominant path are regarded as the same\cite{AC_FHlawatsch_WirelessRapidly}), the delay-Doppler-angle channel $\mathcal{H}$ is sparse along the delay dimension $\ell$. Assuming that the largest path delay is $\tau_{\rm max}$, then $\mathcal{H}$ has finite support $\left[ 0:M_{\rm max}-1\right]$ along the delay dimension $\ell$, where $M_{\rm max}\approx\tau_{\rm max}M\Delta f$.
	
	Additionally, the Doppler frequency of the $s_i$-th subpath in the $i$-th dominant path can be expressed as $\nu_{s_i}=\frac{v}{\lambda}\sin\phi_{s_i}$\cite{AC_FHlawatsch_WirelessRapidly}, where $v$ is the moving velocity of the user and $\phi_{s_i}$ is the
	angle between the user’s moving direction and the arriving direction of the $s_i$-th subpath. Therefore, the maximum Doppler of a subpath is $\frac{\nu_{\rm max}}{2}=\frac{v}{\lambda}$. Since $\phi_{s_i}$ is distributed in $[-\pi/2,\pi/2)$, $\nu_{s_i}$ is distributed in $[-\frac{\nu_{\rm max}}{2},\frac{\nu_{\rm max}}{2})$. Therefore, $\mathcal{H}$ has finite support $\left[ -\frac{N_{\rm max}}{2}:\frac{N_{\rm max}}{2}-1\right]$ along the Doppler dimension $k$, where $N_{\rm max}\approx\nu_{\rm max}NT$. For example, for the typical subcarrier spacing $\Delta f=15$ kHz and carrier frequency 2.15 GHz, the maximum Doppler of a user with a speed of 180 km/h equals to $\frac{\nu_{\rm max}}{2}=358$Hz. Thus, $\frac{N_{\rm max}}{2}\approx\frac{\nu_{\rm max}}{2}NT\approx\frac{\nu_{\rm max}}{2}N/\Delta f\approx 0.05\frac{N}{2}$. There are only about 5\% dominant elements along the Doppler dimension. That is to say, the delay-Doppler-angle channel $\mathcal{H}$ is block-sparse along the Doppler dimension $k$, where the unique non-zero block is centered around $k=0$ but the length of the non-zero block is unknown.
	
	Finally, for the angle dimension $r$, since the angle spread of a dominant path is small, $\psi_{s_i}$ is distributed in $N_{\rm p}$ pieces in $[-1/2,1/2)$. Therefore, the delay-Doppler-angle channel $\mathcal{H}$ is burst-parse \cite{TWC_ALiu_BurstSparsity} along the angle dimension $r$. There are $N_{\rm p}$ non-zero blocks but the start position of each block is unknown, since the path may arrive from any directions. Note that the difference between the burst-sparsity and the traditional block-sparsity is that, the start position of the non-zero burst is not necessarily to be $\{1, 1+D,1+2D, \cdots\}$ where $D$ is the length of non-zero blocks.
	
	To sum up, we decompose the multipaths of time-variant channels to show its structured sparsity along the delay dimension, Doppler dimension, and angle dimension as shown in Fig. \ref{Fig_DDA_Channel}. The 3D channel tensor $\mathcal{H}$ is sparse along the delay dimension, block-sparse along the Doppler
	dimension, and burst-sparse along the angle dimension. This 3D structured sparsity can be used to estimate the CSI with low pilot overhead.

	\subsection{Formulation of Downlink Channel Estimation}\label{S4.2}
\begin{figure}[t] 
		\vspace{-0mm}
		\center{\includegraphics[width=0.9\columnwidth]{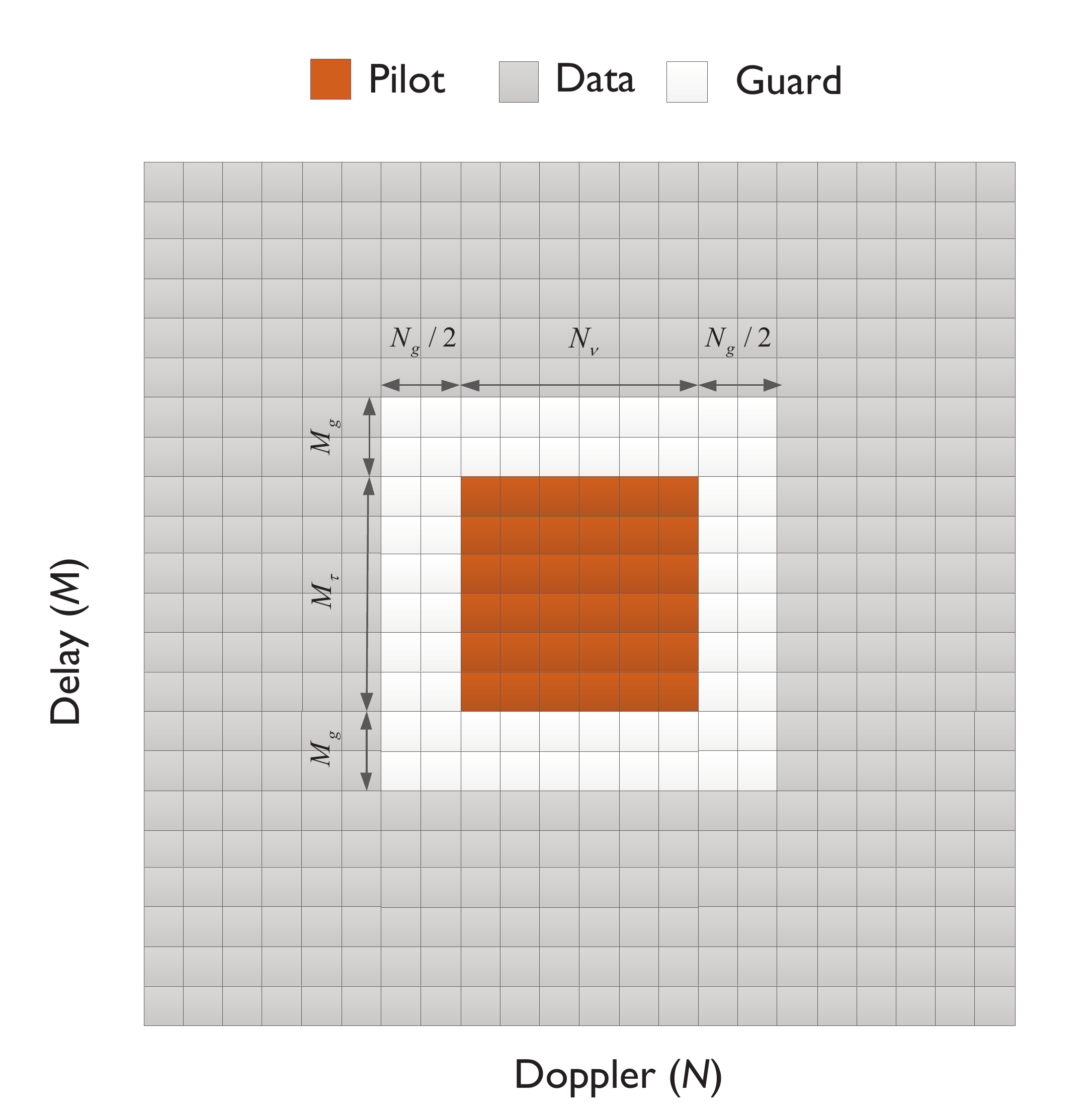}}
		\vspace{-0mm}
		\caption{An OTFS frame in the delay-Doppler domain with pilots and guard intervals.}
		\label{Fig_PilotPattern}
		\vspace{-0mm}
	\end{figure}

	Fig. \ref{Fig_PilotPattern} shows an OTFS frame of size $M\times N$ in the delay-Doppler domain. The length of pilots along the Doppler dimension and the delay dimension are $N_\nu$ and $M_\tau$, satisfying that $ N_\nu\ge N_{\rm max}$ and $M_\tau\ge M_{\rm max}$. We propose to use complex Gaussian random sequences as the training pilots. To avoid interference between pilots and data caused by the two-dimensional periodic convolution in the delay-Doppler domain, guard intervals are required. Note that the delay-Doppler-angle channel $\mathcal{H}$ in OTFS massive MIMO systems has finite supports $\left[ -\frac{N_{\rm max}}{2}:\frac{N_{\rm max}}{2}-1\right]$ along the Doppler dimension and $[0:M_{\rm max}-1]$ along the delay dimension. The length of guard intervals should be $\frac{N_{\rm g}}{2}\ge\frac{N_{\rm max}}{2}-1$ along the Doppler dimension and $M_{\rm g}\ge M_{\rm max}-1$ along the delay dimension as shown in Fig. \ref{Fig_PilotPattern}. To reduce the overall pilot overhead in OTFS massive MIMO systems, we propose the non-orthogonal pilot pattern, i.e., the transmit pilots at different antennas are completely overlapped in the delay-Doppler domain, but the complex Gaussian random sequences (pilots) at different antennas are independent. 
	
	The training pilots in the delay-Doppler domain at the $(p+1)$-th antenna are denoted as $x_{\ell,k,p}$ with $\ell=0,1,\cdots, M_\tau-1$, $k=-\frac{N_\nu}{2},\cdots,0,\cdots,\frac{N_\nu}{2}-1$, and $p=0,1,\cdots,N_{\rm t}-1$. The OTFS frames at $N_{\rm t}$ antennas will be modulated and transmitted simultaneously. After passing the channel, the received signal is demodulated, and then the guard intervals are discarded. According to (\ref{eq_YDDlk2}), the received pilots in the delay-Doppler domain at the user side can be expressed as
	\begin{align}\label{eq_ykl1} 
	y_{\ell,k}=\sum_{p=0}^{N_{\rm t}-1}\sum_{\ell'=0}^{M_{\rm g}-1} \sum_{k'=-\frac{N_{\rm g}}{2}}^{\frac{N_{\rm g}}{2}-1}w_{\ell-\ell',k'}H^{\rm DDS}_{\ell',k',p}x_{\ell-\ell',k-k',p}+v_{\ell,k},
	\end{align}
	where $w_{\ell-\ell',k'}=e^{j2\pi\frac{(\ell-\ell') k'}{N(M+N_{\rm CP})}}$ is the compensate phase, $k=-\frac{N_\nu}{2},\cdots,0,\cdots,\frac{N_\nu}{2}-1$, $\ell=0,1,\cdots, M_\tau-1$. The delay-Doppler-space channel $H^{\rm DDS}_{\ell,k,p}$ can be expressed as the DFT of the delay-Doppler-angle channel $H^{\rm DDA}_{\ell,k,r}$ based on (\ref{eq_hklr1}), i.e.,
	\begin{align}\label{eq_sklp2} 
	H^{\rm DDS}_{\ell,k,p}= \sum_{r=-\frac{N_{\rm t}}{2}}^{\frac{N_{\rm t}}{2}-1}H^{\rm DDA}_{\ell,k,r}e^{-j2\pi \frac{rp}{N_{\rm t}}}.
	\end{align}
	By substituting (\ref{eq_sklp2}) into (\ref{eq_ykl1}) and expressing $z_{\ell-\ell',k-k',r}=\sum_{p=0}^{N_{\rm t}-1}e^{-j2\pi \frac{rp}{N_{\rm t}}}x_{\ell-\ell',k-k',p}$, we have 
	\begin{align}\label{eq_ykl2} 
	y_{\ell,k}=\sum_{r=-\frac{N_{\rm t}}{2}}^{\frac{N_{\rm t}}{2}-1}\sum_{\ell'=0}^{M_{\rm g}-1}\sum_{k'=-\frac{N_{\rm g}}{2}}^{\frac{N_{\rm g}}{2}-1}w_{\ell-\ell',k'}H^{\rm DDA}_{\ell',k',r}z_{\ell-\ell',k-k',r}+v_{\ell,k}.
	\end{align}
	
	To simplify the expression, we rewrite (\ref{eq_ykl2}) into the vector-matrix form. We arrange $y_{\ell,k}$ ($k=-\frac{N_\nu}{2},\cdots,0,\cdots,\frac{N_\nu}{2}-1$, $\ell=0,1,\cdots, M_\tau-1$) into column vectors $\mathbf{y}\in \mathbb{C}^{M_\tau N_\nu \times 1}$, where the $(\ell N_\nu+k+N_\nu/2+1)$-th elements of $\mathbf{y}$ equal to $y_{\ell,k}$. We also arrange $H^{\rm DDA}_{\ell',k',r}$ ($k'=-N_{\rm g}/2,\cdots,0,\cdots,N_{\rm g}/2-1,\ell'=0,1,\cdots, M_{\rm g}-1$) into column vector $\mathbf{h}_{r}\in \mathbb{C}^{M_{\rm g}N_{\rm g} \times 1}$, where the $(\ell'N_{\rm g}+k'+{N_{\rm g}/2}+1)$-th elements of $\mathbf{h}_r$ equal to $H^{\rm DDA}_{\ell',k',r}$. 
	As a result, (\ref{eq_ykl2}) can be rewritten in the vector-matrix form as
	\begin{align}\label{eq_y1} 
	\mathbf{y}=\sum_{r=-\frac{N_{\rm t}}{2}}^{\frac{N_{\rm t}}{2}-1}\mathbf{W}\odot\mathbf{Z}_{{\rm c},r}\mathbf{h}_{r} +\mathbf{v},
	\end{align}
	where $\mathbf{Z}_{{\rm c},r}\in \mathbb{C}^{M_\tau N_\nu \times M_{\rm g} N_{\rm g} }$ is the two-dimensional periodic convolution matrix with the $( \ell N_\nu+k+N_\nu/2+1,\ell' N_{\rm g}+k'+N_{\rm g}/2+1) $-th element of $\mathbf{Z}_{{\rm c},r}$ being equal to $z_{\ell-\ell',k-k',r}$, where $k=-\frac{N_\nu}{2},\cdots,0,\cdots,\frac{N_\nu}{2}-1$, $\ell=0,1,\cdots, M_\tau-1$, $k'=-\frac{N_{\rm g}}{2},\cdots,0,\cdots,\frac{N_{\rm g}}{2}-1$, and $\ell'=0,1,\cdots, M_{\rm g}-1$. $\mathbf{W}\in\mathbb{C}^{M_\tau N_\nu \times M_{\rm g} N_{\rm g}}$ is a matrix with the $(\ell N_\nu+k+N_\nu/2+1,\ell'N_{\rm g}+k'+{N_{\rm g}/2}+1)$-th element being $w_{\ell-\ell',k'}$. By denoting $\mathbf{Z}_{\rm c,W}=\left[ \mathbf{W}\odot\mathbf{Z}_{{\rm c},-\frac{N_{\rm t}}{2}},\cdots,\mathbf{W}\odot\mathbf{Z}_{{\rm c},0},\cdots,\mathbf{W}\odot\mathbf{Z}_{{\rm c},\frac{N_{\rm t}}{2}-1}\right]\in\mathbb{C}^{M_\tau N_\nu \times M_{\rm g} N_{\rm g}  N_{\rm t}}$ and $\mathbf{h}=\left[ \mathbf{h}_{-\frac{N_{\rm t}}{2}}^{\rm T},\cdots,\mathbf{h}_{0}^{\rm T},\cdots,\mathbf{h}_{\frac{N_{\rm t}}{2}-1}^{\rm T}\right] ^{\rm T}\in \mathbb{C}^{M_{\rm g} N_{\rm g}  N_{\rm t}}\times 1$, (\ref{eq_y1}) can be expressed as
	\begin{align}\label{eq_y2} 
	\mathbf{y}=\mathbf{Z}_{\rm c,W}\mathbf{h} +\mathbf{v}.
	\end{align}
	
	Note that $\mathbf{h}$ can be inversely vectorized to obtain a truncated delay-Doppler-angle channel $\mathcal{H}_{\rm g}\in\mathbb{C}^{M_{\rm g}\times N_{\rm g} \times N_{\rm t}}$, i.e., $\mathcal{H}_{\rm g}={\rm invec}\{\mathbf{h}\}$, which is composed of the non-zero part of $\mathcal{H}$ with $\ell=0,1,\cdots,M_{\rm g}-1$, $k=-\frac{N_{\rm g}}{2},\cdots,0,\cdots, \frac{N_{\rm g}}{2}-1$, and $r=-\frac{N_{\rm t}}{2},\cdots,0,\cdots, \frac{N_{\rm t}}{2}-1$. In this way, we formulate the OTFS channel estimation problem as a sparse signal recovery problem with the sensing matrix $\mathbf{\Psi}=\mathbf{Z}_{\rm c,W}$
	\begin{align}\label{eq_y3}  \mathbf{y}=\mathbf{\Psi}\mathbf{h}+\mathbf{v}.
	\end{align}
	This problem can be solved by traditional CS algorithms such as the OMP algorithm\cite{JSAC_LLDai_SpectrallyEddicient}. In the next subsection, we propose a 3D-SOMP algorithm to recover the channel vector $\mathbf{h}$ (or the truncated 3D channel $\mathcal{H}_{\rm g}$) in (\ref{eq_y3}) with improved performance compared with the traditional OMP algorithm.
	
	\subsection{3D-SOMP Algorithm}\label{S4.3}
	\begin{algorithm}[th!]
		\renewcommand{\algorithmicrequire}{\textbf{Input:}}
		\renewcommand\algorithmicensure {\textbf{Output:} }
		\caption{Proposed 3D-SOMP Algorithm}
		\label{alg_3D_OMP}
		\begin{algorithmic}[1]
			\STATE\textbf{Input:} \\
			1) Measurements $\mathbf{y}$; 2) Sensing matrix $\mathbf{\Psi}$\\
			\STATE\textbf{Initialization:} \\
			$i=0$  \\
			$\Omega=\emptyset$ \\
			$\mathbf{h}^{(i)}=\textbf{0}$ \quad\quad\% Initialize the channel vector \\
			$\mathbf{r}=\mathbf{y}-\mathbf{\Psi}\mathbf{h}^{(i)}$ \quad\quad\% Initialize the residual measurements \\
			\FOR {$i\le N_{\rm p}$}
			\STATE $i= i+1$
			\STATE $\mathbf{e}=\mathbf{\Psi}^{\rm H}\mathbf{r}$ 
			\STATE $\mathcal{E}={\rm invec}\{\mathbf{e}\}$ 
			\STATE $e_{\tau}(m)=\|\mathbf{E}_{(1)}(m,:)\|$
			\STATE $m_\tau^{(i)}={\rm arg~max}_m e_\tau(m)$   \quad\quad\% Delay-dimension support
			\STATE $e_\nu(n)=\|\mathcal{E}(m_\tau^{(i)},n,:)\|$ 
			\STATE $n_\nu^{(i)}={\rm arg~min}_n \left\|\mathbf{e}_\nu\left( \frac{N_{\rm g}}{2}-n:\frac{N_{\rm g}}{2}+n-1\right) \right\|$, s.t. $\|\mathbf{e}_\nu\left( \frac{N_{\rm t}}{2}-n:\frac{N_{\rm t}}{2}+n-1\right) \left\| \ge\epsilon\|\mathbf{e}_\nu\right\|$
			\STATE $\Lambda_\nu^{(i)}=\left\{\frac{N_{\rm g}}{2}-n_\nu^{(i)},\cdots,\frac{N_{\rm g}}{2},\cdots,\frac{N_{\rm g}}{2}+n_\nu^{(i)}-1\right\}$ \quad\quad\% Doppler-dimension support 
			\STATE $e_\theta(r)=\left \| \mathcal{E}\left( m_\tau^{(i)},\Lambda_\nu^{(i)},r\right)\right\|$
			\STATE $\mathbf{d}_\theta = \mathbf{L}^{\rm H}\mathbf{e}_\theta$ \quad\quad\% Lifting transformation
			\STATE $g_\theta(r) = \left\| \mathbf{D}_\theta(r,:)\right\| $
			\STATE $p_s={\rm arg~max}_r g_\theta(r)$  \quad\quad\% Start position of the non-zero burst
			\STATE $\Lambda_\theta^{(i)}=\{p_s,p_s+1,\cdots,p_s+D-1\}$  \quad\quad\% Angle-dimension support 
			\STATE $\Omega=\Omega\cup(m_\tau^{(i)},\Lambda_\nu^{(i)},\Lambda_\theta^{(i)})$ \quad\% delay-Doppler-angle 3D support
			\STATE $\mathbf{h}^{(i)}|_\Omega=\mathbf{\Psi}_\Omega^\dagger\mathbf{y}$, $\mathbf{h}^{(i)}|_{\Omega^c}=0$  \quad\quad\% Partial channel estimate 
			\STATE $\mathbf{r}=\mathbf{y}-\mathbf{\Psi}\mathbf{h}^{(i)}$  
			\ENDFOR
			\STATE\textbf{Output:} \\
			Recovered channel vector $\mathbf{\hat{h}}=\mathbf{h}^{(N_{\rm p})}$.\\
		\end{algorithmic}
	\end{algorithm}

	The proposed 3D-SOMP algorithm is presented in \textbf{Algorithm \ref{alg_3D_OMP}}. We borrow the main idea of OMP to obtain the correlation vector $\mathbf{e}$ between the columns of sensing matrix $\mathbf{\Psi}$ and the residual measurements $\mathbf{r}=\mathbf{y}-\mathbf{\Psi}\mathbf{h}^{(0)}$ with $\mathbf{h}^{(0)}=\mathbf{0}$ being the initial channel estimate
	\begin{align}\label{eq_e}  
	\mathbf{e}=\mathbf{\Psi}^{\rm H}\mathbf{r}.
	\end{align}
	For the traditional OMP algorithm, the support of the sparse channel vector $\mathbf{h}$ can be identified by finding the columns of $\mathbf{\Psi}$ that is most correlated to the residual measurement $\mathbf{r}$. Different from OMP, to use the 3D structured sparsity of $\mathbf{h}$ (or $\mathcal{H}_{\rm g}$), we rearrange the correlation vector $\mathbf{e}$ as a tensor $\mathcal{E}\in\mathbb{C}^{M_{\rm g}\times N_{\rm g} \times N_{\rm t}}$ in step 6,
	\begin{align}\label{eq_E}  
	\mathcal{E}={\rm invec}\{\mathbf{e}\}.
	\end{align}
	
	For the sake of presentation, we first introduce some notations of a $N$-dimensional ($N\ge 3$) tensor $\mathcal{M}\in\mathbb{C}^{I_1\times I_2 \times,\cdots,\times I_N}$. The mode-$n$ fiber is obtained by fixing all indexes but the $n$-th index of $\mathcal{M}$, i.e., $\mathcal{M}(i_1,i_2,\cdots,i_{n-1},:,i_{n+1},\cdots,i_N)$. The slice is obtained by fixing all but two indexes of $\mathcal{M}$, i.e., $\mathcal{M}(i_1,i_2,\cdots,i_{n-1},:,:,i_{n+2},\cdots,i_N)$. Finally, the unfolding operation transforms a $N$-dimensional tensor to a 2D matrix. The mode-$n$ unfolding matrix $\mathbf{M}_{(n)}\in\mathbb{C}^{I_n\times I_1I_2\cdots I_{n-1}I_{n+1}\cdots I_N}$ can be obtained by arranging all the mode-$n$ fibers as the columns of $\mathbf{M}_{(n)}$.
	
	Our proposed 3D-SOMP algorithm identifies the 3D support of each dominant path in an one-by-one fashion. For each dominant path, the algorithm starts by obtaining the mode-1 unfolding matrix $\mathbf{E}_{(1)}\in\mathbb{C}^{M_{\rm g}\times  N_{\rm g} N_{\rm t} }$. By calculating the $\ell_2$-norm of row vectors of $\mathbf{E}_{(1)}$, the correlation vector $\mathbf{e}_\tau\in\mathbb{C}^{M_{\rm g}\times1}$ along the delay dimension is obtained with the $m$-th element
	\begin{align}\label{eq_etau}  
	e_{\tau}(m)=\|\mathbf{E}_{(1)}(m,:)\|.
	\end{align}
	Thus, the delay-dimension index $m_\tau^{(i)}$ of the $i$-th dominant path can be obtain by finding the largest element of $\mathbf{e}_\tau$, i.e., $m_\tau^{(i)}={\rm arg~max}_m e_\tau(m)$.
	
	Then, the user fixes the delay-dimension index $m_\tau^{(i)}$ and focuses on the slice $\mathcal{E}(m_\tau^{(i)},:,:)\in\mathbb{C}^{N_{\rm g} \times N_{\rm t} }$ to identify the Doppler- and angle-dimension support. By calculating the $\ell_2$-norm of row vectors of the slice $\mathcal{E}(m_\tau^{(i)},:,:)$, the correlation vector $\mathbf{e}_\nu\in\mathbb{C}^{N_{\rm g}\times1}$ along the Doppler dimension is obtained with the $n$-th element 
	\begin{align}\label{eq_enu}  
	e_\nu(n)=\|\mathcal{E}(m_\tau^{(i)},n,:)\|.
	\end{align}
	Since the truncated 3D channel $\mathcal{H}_{\rm g}$ is block-sparse along the Doppler dimension and there is only one non-zero block centered around $\nu=0$, only the length of the non-zero block is unknown. It can be estimated by finding a smallest block in the Doppler-dimension correlation vector $\mathbf{e}_\nu$, where the ratio between the block's norm and $\|\mathbf{e}_\nu\|$ should be larger than a threshold $\epsilon$, i.e., 
	\begin{align}\label{eq_nnu}  
	n_\nu^{(i)}={\rm arg~min}_n \left\|\mathbf{e}_\nu\left( \frac{N_{\rm g}}{2}-n:\frac{N_{\rm g}}{2}+n-1\right) \right\|, \\\nonumber
	s.t. \left\|\mathbf{e}_\nu\left( \frac{N_{\rm g}}{2}-n:\frac{N_{\rm g}}{2}+n-1\right)\right \| \ge\epsilon\|\mathbf{e}_\nu\|.
	\end{align}
	Thus, the Doppler-dimension support of the $i$-th dominant path is obtained as $\Lambda_\nu^{(i)}$ in step 11. 
	
	Finally, we focus on $\mathcal{E}\left( \Lambda_\nu^{(i)},m_\tau^{(i)},:\right) $ to obtain the angle-dimension support of the $i$-th dominant path. Similarly, by calculating the $\ell_2$-norm of column vector of $\mathcal{E}\left( \Lambda_\nu^{(i)},m_\tau^{(i)},:\right) $, the angle-dimension correlation vector $\mathbf{e}_\theta\in\mathbb{C}^{N_{\rm t}\times1}$ is obtained with the $r$-th element 
	\begin{align}\label{eq_etheta}  
	e_\theta(r)=\left \| \mathcal{E}\left( \Lambda_\nu^{(i)},m_\tau^{(i)},r\right)\right\|.
	\end{align}
	As we have discussed in the previous subsection, the truncated 3D channel $\mathcal{H}_{\rm g}$ is burst-sparse along the angle dimension. The length of the non-zero burst is assumed as $D$. The user needs to estimate the start position of the non-zero burst which is correlated with the AoD of the $i$-th
	dominant path. The user first transforms the burst sparsity into the traditional block sparsity through a lifting transformation method following \cite{TWC_ALiu_BurstSparsity}. In this method, a burst-sparse vector of size $N_{\rm t}\times 1$ is connected to a block-sparse vector with a higher diemnsion $N_{\rm t}D\times 1$ via a lifting matrix $\mathbf{L}\in\{0,1\}^{N_{\rm t}\times N_{\rm t}D}$. The start position of the non-zero burst in the burst-sparse vector is correlated with the support of the non-zero block in the higher-dimensional block-sparse vector. The $((i-1)D+j)$-th column of $\mathbf{L}$ ($i=1,2,\cdots,N_{\rm t}$ and $j=1,2,\cdots,D$) only has one non-zero
	element 1 at location $i\oplus j$ where
	\begin{align}\label{eq_ij} 
	i\oplus j= \left\{ \begin{array}{l}
	i + j,\quad\quad\quad\,\,{{\rm if}}\, i + j \le {N_{\rm t}},\\
	i + j - {N_{\rm t}},\quad {{\rm if}}\,i + j> {N_{\rm t}}.
	\end{array} \right. 
	\end{align}
	To transform the burst sparsity of the truncated 3D channel $\mathcal{H}_{\rm g}$ along the angle dimension into the traditional block sparsity, the angle-dimension correlation vector $\mathbf{e}_\theta$ is modified by the lifting matrix $\mathbf{L}$ as
	\begin{align}\label{eq_dtheta}  
	\mathbf{d}_\theta = \mathbf{L}^{\rm H}\mathbf{e}_\theta.
	\end{align} 
	Then $\mathbf{d}_\theta\in \mathbb{C}^{N_{\rm t}D\times 1}$ is rearranged as a $N_{\rm t}\times D$ matrix $\mathbf{D}_\theta$. By calculating the $\ell_2$-norm of the row vectors of $\mathbf{D}_\theta$, we obtain $\mathbf{g}_\theta\in \mathbb{C}^{N_{\rm t}\times 1}$ in step 14. Thus, the start position $p_s$ of the non-zero burst is obtained by finding the largest element of $\mathbf{g}_\theta$. Therefore, the angle-dimension support correlated to the $i$-th dominant path can be obtained as $\Lambda_\theta^{(i)}=\{p_s,p_s+1,\cdots,p_s+D-1\}$ in step 16.
	
	Up to this point, the delay-Doppler-angle 3D support in the $i$-th iteration can be obtained as $\Omega=\Omega\cup\left( m_\tau^{(i)},\Lambda_\nu^{(i)},\Lambda_\theta^{(i)}\right) $. The user can partially estimate the channel through the LS as $\mathbf{h}^{(i)}|_\Omega=\mathbf{\Psi}_\Omega^\dagger\mathbf{y}$, $\mathbf{h}^{(i)}|_{\Omega^c}=0$, where $\Omega^c$ denotes the complementary set of $\Omega$. Then, the residual measurements is computed by subtracting the contribution of $\mathbf{h}^{(i)}$ in the $i$-th iteration in step 19.
	After $N_{\rm p}$ iterations, the complete channel estimate is obtained as $\mathbf{\hat{h}}=\mathbf{h}^{(N_{\rm p})}$.

	\subsection{Performance Comparison}\label{S4.4}
	For the traditional impulse based channel estimation technique (extended to OTFS massive MIMO systems), the pilot overhead is $\propto N_{\rm t}N_{\rm max}M_{\rm max}$. In our proposed channel estimation technique, the pilot overhead (i.e., the length of measurements) is $\propto S\log(L)$, where $S$ and $L$ are the sparsity level and length of the sparse vector $\mathbf{h}$, according to CS theory\cite{TIT_Donoho_CS}. For our problem formulation in the last subsection, $S=N_{\rm max}N_{\rm p}D$ and $L=N_{\rm g}M_{\rm g}N_{\rm t}$. Therefore, the pilot overhead of our proposed channel estimation technique is $\propto N_{\rm max}N_{\rm p}D\log(N_{\rm g}M_{\rm g}N_{\rm t})$. Note that the number of dominant paths is usually small, e.g., $N_{\rm p}=6$\cite{3GPPTR_SCM}. Since the angle spread of a dominant path is usually not large, the length of non-zero block along the angle dimension $D$ is usually much smaller than the number of BS antennas $N_{\rm t}$, e.g., $D\approx N_{\rm t}/10$\cite{3GPPTR_SCM}. The lengths of guard intervals $N_{\rm g}$ and $M_{\rm g}$ can be set as $N_{\rm max}$ and $M_{\rm max}$. Therefore, the pilot overhead of the proposed 3D-SOMP based channel estimation is much lower than that of the previously proposed impulse based channel estimation.
	\section{Simulation Results}\label{S5}
	In this section, we investigate the performance of the proposed 3D-SOMP based channel estimation technique, in terms of the normalized mean square error (NMSE) of channel estimation. The traditional impulse based channel estimation technique is presented as a benchmark, where we use the LS estimator to estimate the delay-Doppler channel $H^{\rm DD}_{\ell,k}$ ($k =-\frac{N}{2} ,\cdots,0,\cdots,\frac{N}{2} - 1$ and $\ell  =0,1,\cdots,M - 1$) of each antenna from (\ref{eq_Ykl4}) as
	\begin{align}
\hat H_{\ell ,k}^{{\rm{DD}}} = \left\{ {\begin{array}{*{20}{l}}
	{Y_{\ell ,k}^{{\rm{DD}}}{e^{ - j2\pi \frac{{\ell k}}{{N(M + {N_{{\rm{CP}}}})}}}},\quad \begin{array}{*{20}{c}}
		{k \in \left[ { - \frac{{{N_{{\rm{max}}}}}}{2},\frac{{{N_{{\rm{max}}}}}}{2} - 1} \right]}\\
		{\ell  \in \left[ {0,{M_{{\rm{max}}}} - 1} \right]}
		\end{array},}\\
	{0,\quad \quad \quad \begin{array}{*{20}{c}}
		{k \notin \left[ { - \frac{{{N_{{\rm{max}}}}}}{2},\frac{{{N_{{\rm{max}}}}}}{2} - 1} \right]}\\
		{\ell  \notin \left[ {0,{M_{{\rm{max}}}} - 1} \right]}
		\end{array}.}
	\end{array}} \right.
	\end{align}
	The NMSE of the traditional impulse based channel estimation technique is computed as
	\begin{align}
	{\rm NMSE}= \frac{\sum_{k=-\frac{N}{2}}^{k=\frac{N}{2}}\sum_{\ell=0}^{\ell=M-1}|\hat{H}^{\rm DD}_{\ell,k}-H^{\rm DD}_{\ell,k}|^2}{\sum_{k=-\frac{N}{2}}^{k=\frac{N}{2}}\sum_{\ell=0}^{\ell=M-1}|H^{\rm DD}_{\ell,k}|^2},
	\end{align}
	which will be averaged over $N_{\rm t}$ antennas. For the proposed channel estimation technique, the channel vector $\mathbf{h}$ in (\ref{eq_y3}) can be estimated through the proposed 3D-SOMP algorithm as $\mathbf{\hat{h}}$. Then, $\mathbf{\hat{h}}$ is rearranged as a $M_{\rm g}\times N_{\rm g} \times N_{\rm t}$ tensor $\mathcal{\hat{H}}_{\rm g}={\rm invec}\{\mathbf{\hat{h}}\}$. Thus, the delay-Doppler-angle channel can be estimated as $\mathcal{\hat{H}}|_\Gamma=\mathcal{\hat{H}}_{\rm g}$ and $\mathcal{\hat{H}}|_{\Gamma^c=0}$, where $\Gamma$ is the index set of $\mathcal{\hat{H}}_{\rm g}$.
	The NMSE of the proposed 3D-SOMP based channel estimation technique is computed as 
	\begin{align}
	{\rm NMSE}= \frac{\|\mathcal{\hat{H}}-\mathcal{H}\|^2}{\|\mathcal{H}\|^2}.
	\end{align}
	We also present the NMSE of the traditional OMP based channel estimation technique for comparison when the traditional OMP algorithm is used to recover $\mathbf{h}$ in (\ref{eq_y3}).
	
	We simulate the standardized spatial channel model in 3GPP considering the urban macro cell environment \cite{ScmImplementation}. The detailed system parameters are summarized in Table \ref{Table_1}. We define the pilot overhead ratio $\eta$ as the ratio between the number of resource units for pilot transmission and the number of total resource units in the delay-Doppler domain. We will compare the NMSE performance of the proposed 3D-SOMP based channel estimation technique, the traditional impulse based channel estimation technique, and the traditional OMP based channel estimation technique against the pilot overhead ratio, the number of BS antennas, and the signal-to-noise ratio (SNR).
	\begin{table}[tb!]
		\begin{center}
			\caption{System parameters for simulation}  \label{Table_1}
			\begin{tabular}{|c|c|}
				\hline
				{\textbf{Parameter}} & \textbf{Values} \\
				\hline
				\tabincell{c}{Carrier frequency (GHz) } & 2.15 \\
				\hline
				\tabincell{c}{Duplex mode} & FDD \\
				\hline			
				\tabincell{c}{Subcarrier spacing (kHz)} & 15 \\
				\hline
				\tabincell{c}{Cyclic prefix duration (us)} & 16.6 \\
				\hline
				\tabincell{c}{FFT size} & 1024 \\
				\hline
				\tabincell{c}{Transmission bandwidth ($\#$ of resource blocks)} &  50 \\
				\hline
				\tabincell{c}{Size of a OTFS frame $(M,N)$} &  $(600,12)$ \\
				\hline
				\tabincell{c}{$\#$ of BS antennas} &  $8\sim 64$ \\
				\hline
				\tabincell{c}{$\#$ of user antennas} &  1 \\
				\hline
				\tabincell{c}{Channel model:3GPP standardized channel model} &  Urban macro cell \\
				\hline
				\tabincell{c}{$\#$ of dominant channel paths} &  6 \\
				\hline
				\tabincell{c}{$\#$ of sub-paths per dominant path} &  20 \\
				\hline
				\tabincell{c}{User velocity (km/h)} &  360 \\
				\hline
				
			\end{tabular}
		\end{center}
	\end{table}
	
\begin{figure}[tb!]
		\vspace*{-0mm}
		\begin{center}
			\includegraphics[width=1\columnwidth]{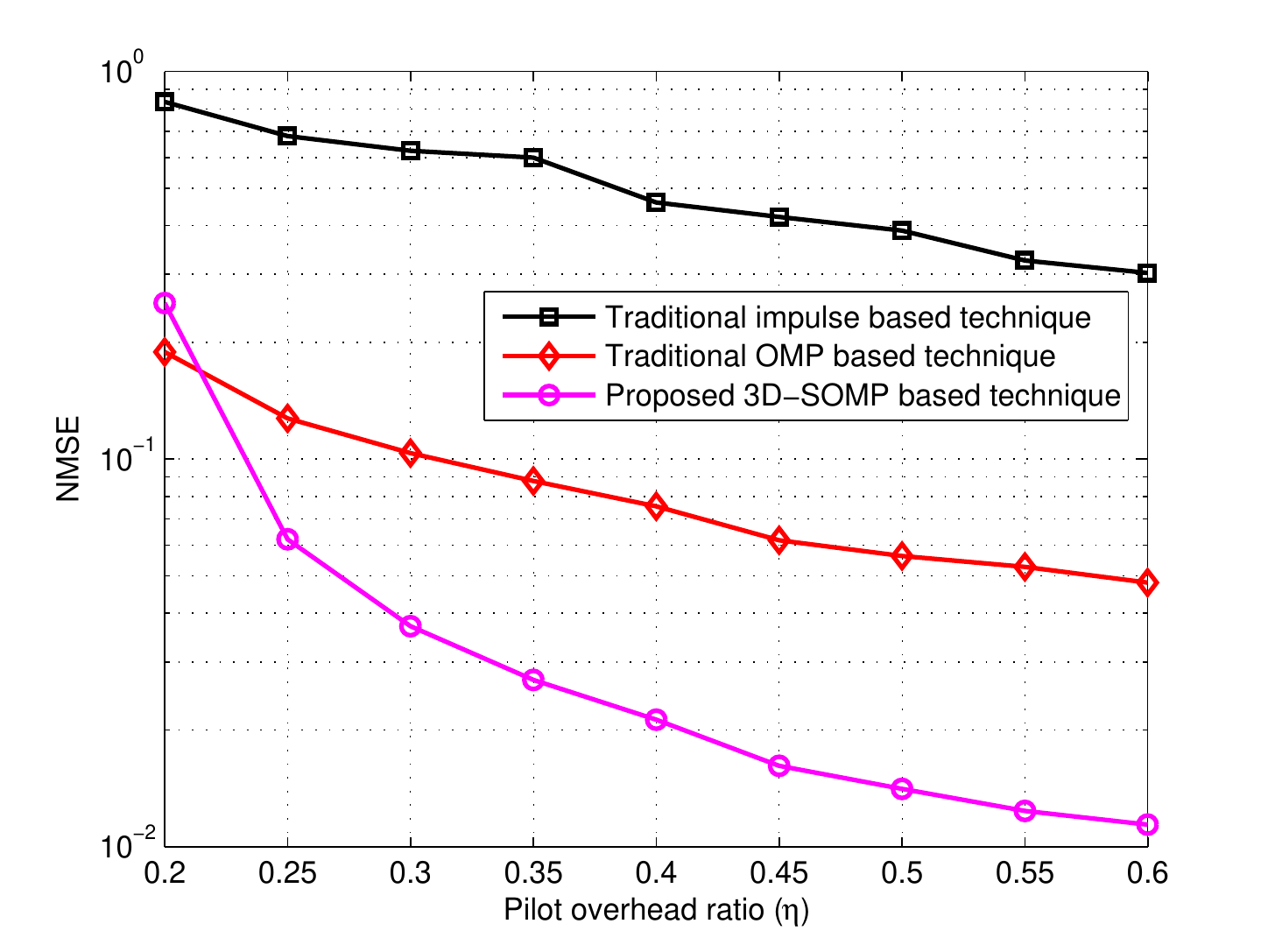}
		\end{center}
		\vspace*{-3mm}
		\caption{The NMSE performance comparison against the pilot overhead ratio $\eta$. The number of BS antennas is 16 and the SNR is 5 dB.}
		\label{fig_Plot_NMSE_eta}
	\end{figure}

	In Fig. \ref{fig_Plot_NMSE_eta}, we show the NMSE performance comparison against the pilot overhead ratio $\eta$. The number of BS antennas is 16 and the SNR is 5 dB. We observe that the proposed 3D-SOMP based channel estimation technique outperforms the traditional impulse based channel estimation technique, when the same pilot overhead ratio is considered. The traditional impulse based technique does not perform well due to insufficient pilot overhead when the pilot overhead ratio is small, i.e, the intervals between two adjacent impulses are smaller than $N_{\rm max}$ along the Doppler dimension and/or smaller than $M_{\rm max}$ along the delay dimension. Therefore, interference from adjacent impulses will degrade the NMSE performance of the traditional impulse based channel estimation technique. By contrast, non-orthogonal pilots are used for the proposed 3D-SOMP based channel estimation technique. The required pilot overhead is $\propto N_{\rm max}N_{\rm p}D\log(N_{\rm g}M_{\rm g}N_{\rm t})$, which is much smaller than that of the traditional impulse based channel estimation technique. For example, to achieve the NMSE of 0.03, only 32\% pilot overhead ratio is required for the proposed 3D-SOMP based channel estimation technique. For the traditional impulse based channel estimation technique, 60\% pilot overhead ratio is required to achieve a NMSE of 0.3. Moreover, the proposed 3D-SOMP based channel estimation technique has better performance than the traditional OMP based channel estimation technique, which results from its use of the 3D structured sparsity of the delay-Doppler-angle channel in OTFS massive MIMO systems. 
	
\begin{figure}[tb!]
		\begin{center}
			\includegraphics[width=1\columnwidth]{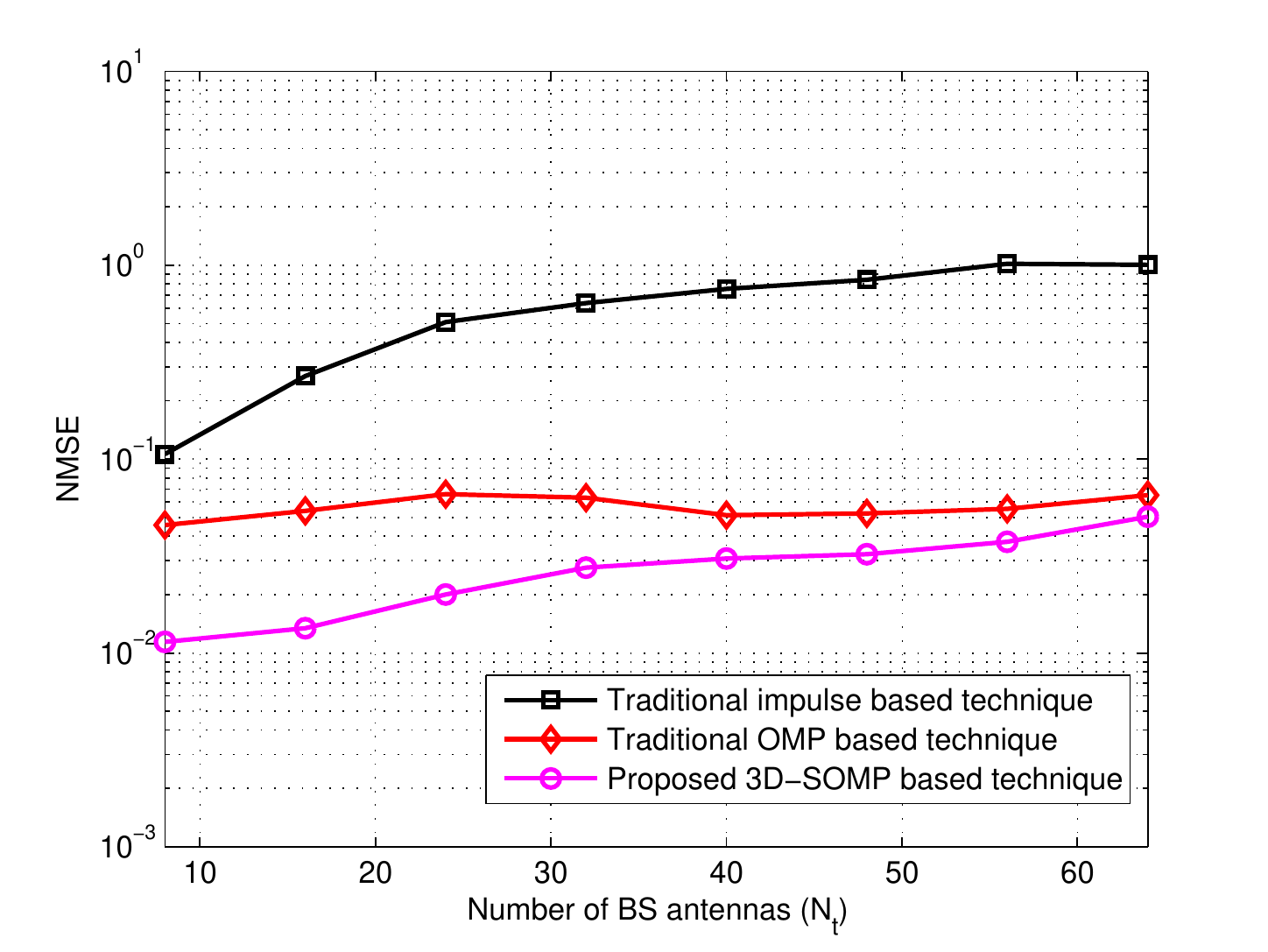}
		\end{center}
		\vspace*{-3mm}
		\caption{The NMSE performance comparison against the number of BS antennas. The pilot overhead ratio is $50\%$ and the SNR is 5 dB.}
		\label{fig_Plot_NMSE_BS}
	\end{figure}

	In Fig. \ref{fig_Plot_NMSE_BS}, we present the NMSE performance comparison against the number of BS antennas $N_{\rm t}$. The pilot overhead ratio is set as $50\%$ and the SNR is 5 dB. We observe that the NMSE performance of the traditional impulse based channel estimation technique severely degrades (NMSE is larger than $10^{-1}$) when the the number of BS antennas increases larger than 8. This is due to the insufficient intervals between two adjacent impulses when the number of BS antennas is large while the pilot overhead ratio is constant. On the contrary, the proposed 3D-SOMP based channel estimation technique works well with a large number of BS antennas. Moreover, the proposed 3D-SOMP based channel estimation technique outperforms the traditional OMP based channel estimation technique in the considered numbers of BS antennas. 
	\begin{figure}[tb!]
		\begin{center}
			\includegraphics[width=1\columnwidth]{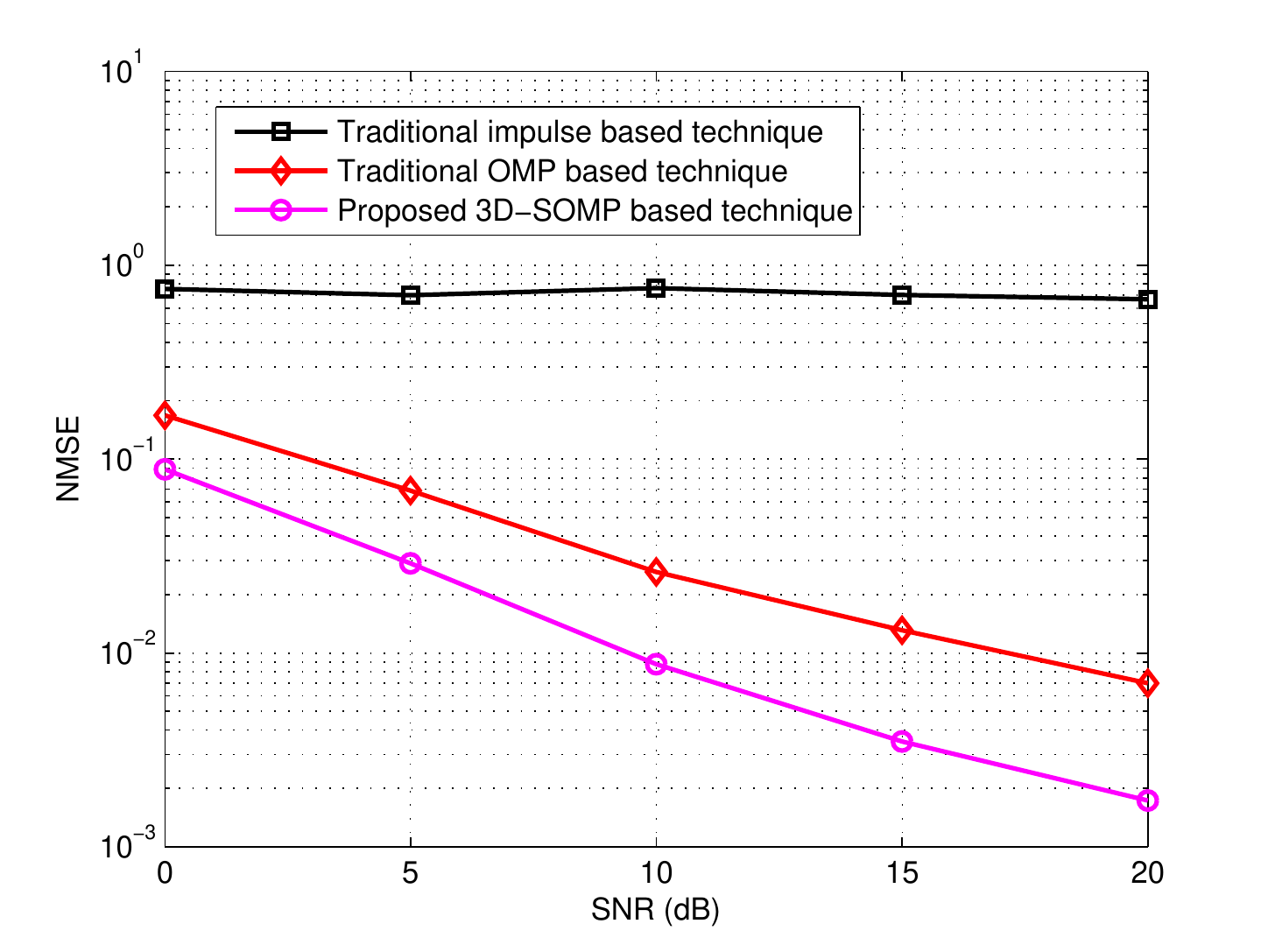}
		\end{center}
		\vspace*{-3mm}
		\caption{The NMSE performance comparison against the SNR. The number fo BS antennas is 32 and the pilot overhead ratio is $50\%$.} \label{fig_NMSE_SNR}
	\end{figure}
	
	In Fig. \ref{fig_NMSE_SNR}, we show the NMSE performance comparison against the SNR. The number of antennas is set as 32 and the pilot overhead ratio is $50\%$. We observe that the proposed 3D-SOMP based channel estimation technique outperforms the traditional impulse based channel estimation technique. The traditional impulse based channel estimation technique has a NMSE floor due to the interference among multiple antennas resulted from the insufficient pilot overhead. For the proposed 3D-SOMP based technique, the NMSE performance is improved with the increased SNR. Moreover, the proposed 3D-SOMP based technique outperforms the traditional OMP based technique by about 6 dB.

	
	\section{Conclusions}\label{S6}
	In this paper, we studied the OTFS modulation for massive MIMO systems for the first time with the focus on channel estimation. Specifically, we transformed the time-variant massive MIMO channels into the delay-Doppler-angle 3D channel in OTFS massive MIMO systems. We found that the 3D channel is structured sparse, i.e., sparse along the delay dimension, block-sparse along the Doppler dimension, and burst-sparse along the angle dimension. Based on the 3D structured sparsity, we formulated the downlink channel estimation problem as a sparse signal recovery problem and solved it with the proposed 3D-SOMP algorithm. Simulation results verified the superior performance of our proposed technique. For future research, we will focus on some open problems in OTFS massive MIMO systems such as the low-complexity equalizer, downlink precoding, and efficient channel feedback.
	
	\section*{Appendix I}\label{S7.1}
	Proof of Lemma 1.
	\begin{IEEEproof}
		Based on the OTFS modulation, each column vector $\mathbf{s}_i\in\mathbb{C}^{M\times 1}$ ($i=1,2,\cdots,N$) of $\mathbf{S}$ is an OFDM symbol (without CP),
		\begin{align}\label{eq_si} 
		\mathbf{s}_i=\mathbf{X}^{\rm DD}\mathbf{f}_i^*,
		\end{align}
		where $\mathbf{X}^{\rm DD}$ is the 2D data block in the delay-Doppler domain and $\mathbf{f}_i\in\mathbb{C}^{N\times 1}$ is the $i$-th column vector of the DFT matrix $\mathbf{F}_{\rm N}$. Then, CP is added to each OFDM symbol and these OFDM symbols with CPs are transmitted by the BS. After passing through the channel, the received OFDM symbols are removed with CPs and arranged in the columns of matrix $\mathbf{Z}=\left[ \mathbf{z}_1,\mathbf{z}_2,\cdots,\mathbf{z}_N\right]\in\mathbb{C}^{M\times N}$,
		\begin{align}\label{eq_Z} 
		\mathbf{Z}=\mathbf{R}_{\rm CP}\mathbf{R}.
		\end{align}
		To avoid the inter-symbol interference, the length of CP $N_{\rm CP}$ is usually larger than the channel length $L$, i.e., $N_{\rm CP}>L$. Thus, the $i$-th received OFDM symbol (without CP) $\mathbf{z}_i\in\mathbb{C}^{M\times 1}$ is given by the circular convolution of the $i$-th transmit OFDM symbol (without CP) $\mathbf{s}_i$ with the time-variant channel, i.e.,
		\begin{align}\label{eq_zi} 
		\mathbf{z}_i=\mathbf{H}^{\rm c}_i\mathbf{s}_i + \mathbf{v}_i,
		\end{align}
		where $\mathbf{H}^{\rm c}_i\in\mathbb{C}^{M\times M}$ is the circular convolution matrix, whose $(a,b)$-th element can be expressed as $h_{(i-1)(M+N_{\rm CP})+a,(a-b)_M}$ ($a=1,2,\cdots,M$ and $b=1,2,\cdots,M$), where $(a-b)_M$ is the remainder after division of $a-b$ by $M$. $\mathbf{v}_i$ is the additive noise vector. By substituting (\ref{eq_si}) into (\ref{eq_zi}),
		\begin{align}\label{eq_zi1} 
		\mathbf{z}_i=\mathbf{H}^{\rm c}_i\mathbf{X}^{\rm DD}\mathbf{f}_i^* + \mathbf{v}_i.
		\end{align}
		
		The received OFDM symbols $\mathbf{Z}$ without CPs are transformed to the 2D data block in the delay-Doppler domain $\mathbf{Y}^{\rm DD}$ as (\ref{eq_YDD1}), i.e., 
		\begin{align}\label{eqa_YDD1} 
		\mathbf{Y}^{\rm DD}=\mathbf{Z}\mathbf{F}_{\rm N}=\left[ \mathbf{z}_1,\mathbf{z}_2,\cdots,\mathbf{z}_N\right]\mathbf{F}_{\rm N}.
		\end{align}
		We rewrite (\ref{eqa_YDD1}) as 
		\begin{align}\label{eq_YDDa2} 
		\mathbf{Y}^{\rm DD}=\sum_{i=1}^{N}\mathbf{z}_i\mathbf{f}_i^{\rm T}.
		\end{align}
		By substituting (\ref{eq_zi1}) into (\ref{eq_YDDa2}), 
		\begin{align}\label{eqa_YDD2} 
		\mathbf{Y}^{\rm DD}=\sum_{i=1}^{N}\mathbf{H}^{\rm c}_i\mathbf{X}^{\rm DD}\mathbf{f}_i^*\mathbf{f}_i^{\rm T} +\mathbf{V}^{\rm DD},
		\end{align}
		where $\mathbf{V}^{\rm DD}=\left[ \mathbf{v}_1,\mathbf{v}_2,\cdots,\mathbf{v}_N\right]\mathbf{F}_{\rm N}$. We denote the $(\ell+1,k+1+N/2)$-th element of $\mathbf{Y}^{\rm DD}$ and $\mathbf{X}^{\rm DD}$ as $Y^{\rm DD}_{\ell,k}$ and $X^{\rm DD}_{\ell,k}$, where $\ell=0,1,\cdots,M-1$ and $k=-N/2,\cdots,0,\cdots,N/2-1$.	Expanding to sum in (\ref{eqa_YDD2}), $Y^{\rm DD}_{\ell,k}$ is given by
		\begin{align}\label{eqa_YDDlk} 
\begin{array}{l}
Y_{\ell ,k}^{{\rm{DD}}} = \sum\limits_{\ell ' = 0}^{M - 1} {\sum\limits_{k' =  - N/2}^{N/2 - 1} {X_{\ell ',k'}^{{\rm{DD}}}} } \sum\limits_{i = 1}^N {{h_{(i - 1)(M + {N_{{\rm{CP}}}}) + \ell  + 1,{{(\ell  - \ell ')}_M}}}} \\
\quad \quad \quad \quad \quad \quad \quad \quad \quad \times{e^{ - j2\pi (i - 1)\frac{{k - k'}}{N}}} + V_{\ell ,k}^{{\rm{DD}}}.
\end{array}
		\end{align}
		
		We define $\Lambda_{\ell,(\ell-\ell')_M,k-k'}\overset{\Delta}{=}\sum_{i=1}^{N}h_{(i-1)(M+N_{\rm CP})+\ell+1,(\ell-\ell')_M}e^{-j2\pi (i-1)\frac{k-k'}{N}}$ and focus on the calculation of $\Lambda_{l,(\ell-\ell')_M,k-k'}$. We first expand the time-variant channels $h_{\kappa,\ell}$ based on the Fourier series as 
		\begin{align}\label{eqa_hkal}
		h_{\kappa,\ell}=\sum_{p=1}^{P}\omega_{p,\ell}e^{j2\pi \frac{\kappa f_{p,\ell}}{N(M+N_{\rm CP})}},
		\end{align}
		where $P$ is the number of frequency component of time-variant channels. $f_{p,\ell}$ is the $p$-th frequency component of the $\ell$-th channel tap. $\omega_{p,\ell}$ is the non-zero coefficient corresponding to $e^{j2\pi \frac{\kappa f_{p,\ell}}{N(M+N_{\rm CP})}}$. 
		Based on (\ref{eqa_hkal}), $\Lambda_{\ell,(\ell-\ell')_M,k-k'}$ is expressed as 
		\begin{align}\label{eqa_lambda}
\begin{array}{l}
{\Lambda _{\ell ,{{(\ell  - \ell ')}_M},k - k'}} = \\
\sum\limits_{i = 1}^N {\sum\limits_{p = 1}^P {{\omega _{p,{{(\ell  - \ell ')}_M}}}} } {e^{j2\pi \frac{{\left( {(i\! -\! 1)(M \!+\! {N_{{\rm{CP}}}}) + \ell  + 1} \right){f_{p,{{(\ell \! -\! \ell ')}_M}}}}}{{N(M + {N_{{\rm{CP}}}})}}}}{e^{ - j2\pi (i\! -\! 1)\frac{{k - k'}}{N}}}\\
=\!\! \sum\limits_{p = 1}^P {{\omega _{p,{{(\ell  - \ell ')}_M}}}} {e^{j2\pi \frac{{(\ell  + 1){f_{p,{{(\ell  - \ell ')}_M}}}}}{{N(M + {N_{{\rm{CP}}}})}}}}\sum\limits_{i = 1}^N {{e^{j2\pi (i \! - \! 1)\frac{{{f_{p,{{(\ell  - \ell ')}_M}}} \! - \left( {k - k'} \right)}}{N}}}} .
\end{array}
		\end{align}
		We define a function $\Upsilon_N(x)\triangleq\sum_{i=1}^{N}e^{j2\pi\frac{x}{N}(i-1)}=\frac{\sin(\pi x)}{\sin(\pi \frac{x}{N})}e^{j\pi \frac{x(N-1)}{N}}$. Then, (\ref{eqa_lambda}) is rewritten as 
		\begin{align}\label{eqa_lambda1}
		\Lambda_{\ell,(\ell-\ell')_M,k-k'}=\sum_{p=1}^{P}& \omega_{p,(\ell-\ell')_M} e^{j2\pi \frac{(\ell+1) f_{p,(\ell-\ell')_M} }{N(M+N_{\rm CP})}} \\\nonumber &\times\Upsilon_N\left( f_{p,(\ell-\ell')_M} -\left( k-k'\right)\right).
		\end{align}
		
		Now we can define the delay-Doppler CIR $H^{\rm DD}_{\ell,k}$ ($k=-N/2,\cdots,0,\cdots,N/2-1$ and $\ell=0,1,\cdots,M-1$) as 
		\begin{align}
		H^{\rm DD}_{\ell,k}&\overset{\Delta}{=}\Lambda_{0,(\ell)_M,k}\\\nonumber&=\sum_{i=1}^{N}h_{(i-1)(M+N_{\rm CP})+1,(\ell)_M}e^{-j2\pi (i-1)\frac{k}{N}}.
		\end{align}
		Then, $H^{\rm DD}_{\ell-\ell',k-k'}$ is given by 
		\begin{align}\label{eqa_lambda2}
		H^{\rm DD}_{\ell-\ell',k-k'}&=\Lambda_{0,(\ell-\ell')_M,k-k'}\\\nonumber &=\sum_{p=1}^{P}\tilde{\omega}_{p,(\ell-\ell')_M} \Upsilon_N\left( f_{p,(\ell-\ell')_M} - (k-k')\right),
		\end{align}
		where $\tilde{\omega}_{p,(\ell-\ell')_M}=\omega_{p,(\ell-\ell')_M} e^{j2\pi \frac{f_{p,(\ell-\ell')_M}}{N(M+N_{\rm CP})}} $. Now we will prove that $\Lambda_{\ell,(\ell-\ell')_M,k-k'}\overset{N\rightarrow\infty}{=}e^{j2\pi\frac{\ell \left( k-k'\right)}{N(M+N_{\rm CP})}}H^{\rm DD}_{\ell-\ell',k-k'}$. Specifically,
		we first calculate 
		\begin{align}\label{eqa_lambda3}
\begin{array}{l}
{\Lambda _{\ell ,{{(\ell  - \ell ')}_M},k - k'}}{e^{ - j2\pi \frac{{\ell \left( {k - k'} \right)}}{{N(M + {N_{{\rm{CP}}}})}}}} = \\
\sum\limits_{p = 1}^P {{{\tilde \omega }_{p,{{(\ell \!  -\! \ell ')}_M}}}} {e^{j2\pi \frac{{\ell \left( {{f_{p,{{(\ell \!  - \!\ell ')}_M}}} \! - \left( {k\! -\! k'} \right)} \right)}}{{N(M + {N_{{\rm{CP}}}})}}}}{\Upsilon _N}\left( {{f_{p,{{(\ell  - \ell ')}_M}}} \!\! - \!\left( {k \!- \!k'} \right)} \right).
\end{array}
		\end{align}
		It is noticed that the function $\Upsilon_N(x)$ has the following characteristic: $\left| \Upsilon_N(x)\right| \rightarrow 0$ when $|x|\gg1$ \cite{TWC_XGao_BeamspaceChannelEstimation}. Thus we conclude that there are $P$ dominant items in (\ref{eqa_lambda3}), which are obtained when $\left| f_{p,(\ell-\ell')_M} -\left( k-k'\right) \right| <1$. Since $0\le \ell\le M-1$, we have 
		\begin{align}\label{eqa_1N}
		e^{j2\pi \frac{\ell \left( f_{p,(\ell-\ell')_M} -\left( k-k'\right)\right) }{N(M+N_{\rm CP})}} \overset{N\rightarrow\infty}{=}1.
		\end{align}
		Therefore, by combining (\ref{eqa_lambda2}), (\ref{eqa_lambda3}), and (\ref{eqa_1N}),
		\begin{align} \label{eqa_lambda4}
		\Lambda_{\ell,(\ell-\ell')_M,k-k'}\overset{N\rightarrow\infty}{=}H^{\rm DD}_{\ell-\ell',k-k'}e^{j2\pi\frac{\ell \left( k-k'\right)}{N(M+N_{\rm CP})}}.
		\end{align}
		Finally, by substituting (\ref{eqa_lambda4}) into (\ref{eqa_YDDlk}), we prove that
		\begin{align}\label{eqa_YDDlk2} 
		Y^{\rm DD}_{\ell,k}\overset{N\rightarrow\infty}{=}\sum_{\ell'=0}^{M-1}\sum_{k'=-N/2}^{N/2-1}&X^{\rm DD}_{\ell',k'}H^{\rm DD}_{\ell-\ell',k-k'}e^{j2\pi\frac{\ell \left( k-k'\right)}{N(M+N_{\rm CP})}}\\\nonumber&+V^{\rm DD}_{\ell,k}.
		\end{align}
	\end{IEEEproof}
	\bibliographystyle{IEEEtran}
	\bibliography{IEEEabrv,Refference}
\end{document}